\documentclass [11pt]{article}
\usepackage{amsmath,amsthm,amsfonts,amscd,eucal,latexsym,amssymb}
\usepackage{epsfig}   
\def\spa{\hskip -3pt}

\newsymbol\bt 1202           
\def\a{\alpha}
\def\b{\beta}
\def\c{\gamma}
\def\d{\delta}

\def\loc{L^1_{loc}(\Sigma,\omega_\Sigma)}

\def\cG{{\ca G}}
\def\cL{{\ca L}}

\def\cH{{\ca H}}
\def\cD{{\ca D}}
\def\cE{{\ca E}}
\def\cO{{\ca O}}
\def\cS{{\ca S}}
\def\cA{{\ca A}}

\def\cC{{\ca C}}
\def\cX{{\ca X}}
\def\cW{{\ca W}}
\def\cK{{\ca K}}

\def\bE{{\mathbb E}} 
\def\bC{{\mathbb C}}           
 
\def\bI{{\mathbb I}}

\def\bN{{\mathbb N}} 
\def\bM{{\mathbb M}} 
 
\def\bR{{\mathbb R}} 
\def\bP{{\mathbb P}}
 \def\bF{{\mathbb F}}
\def\bS{{\mathbb S}}

\def\bZ{{\mathbb Z}} 
 
\newsymbol\rest 1316         
 
\def\gF{{\mathfrak F}} 
 
\def\gH{{\mathfrak H}}

\def\beq{\begin{eqnarray}}
\def\eeq{\end{eqnarray}}

\def\al{\langle}
\def\cl{\rangle}
\newcommand{\ca}[1]{{\cal #1}}         

\def\Si{\Sigma}

\newcounter{proposition}[section]
\newcounter{theorem}[section]
\newcounter{lemma}[section]
\newcounter{definition}[section]
\def\theproposition{\thesection.\arabic{proposition}}
\def\thetheorem{\thesection.\arabic{theorem}}
\def\thelemma{\thesection.\arabic{lemma}}
\def\thedefinition{\thesection.\arabic{definition}}

\def\s #1 {\section{#1}}

\def\ssa #1 {\ifhmode{\par}\fi\refstepcounter{subsection}
  \noindent {\bf\thesubsection}. {\em #1}.\quad
  \addcontentsline{toc}{subsection}{\protect\numberline{\thesubsection} #1}%
  }

\def\ssb #1 {\ifhmode{\par}\fi\refstepcounter{subsection}
  \noindent {\bf\thesubsection.} {\bf #1.}\quad
  \addcontentsline{toc}{subsection}{\protect\numberline{\thesubsection} #1}%
  }

\def\proposizione {\ifhmode{\par}\fi\refstepcounter{proposition}
  \noindent {\bf Proposition \theproposition}. \quad}
\def\teorema {\ifhmode{\par}\fi\refstepcounter{theorem}
  \noindent {\bf Theorem \thetheorem}. \quad}
\def\lemma {\ifhmode{\par}\fi\refstepcounter{lemma}
  \noindent {\bf Lemma \thelemma}. \quad}
\def\definizione {\ifhmode{\par}\fi\refstepcounter{definition}
  \noindent {\bf Definition \thedefinition}. \quad}

\begin{document} 
 
\hfill{\sl Preprint  UTM  674 - Revised version March 2005} 
\par 
\bigskip 
\par 
\rm

 
\par 
\bigskip 
\LARGE 
\noindent 
{\bf Bose-Einstein condensate and Spontaneous Breaking of  Conformal Symmetry
on Killing Horizons} 
\bigskip 
\par 
\rm 
\normalsize 
 
 
\large 
\noindent {\bf Valter Moretti$^{1,2,a}$} and {\bf Nicola Pinamonti$^{1,2,b}$} \\
Department of Mathematics, Faculty of Science,
University of Trento, \\ via Sommarive 14, 
I-38050 Povo (TN), 
Italy\\\par 
\smallskip\noindent 
$^1$  Istituto Nazionale di Alta Matematica ``F.Severi'',  unit\`a locale  di Trento\\ 
$^2$  Istituto Nazionale di Fisica Nucleare,  Gruppo Collegato di Trento\\
$^a$ E-mail: moretti@science.unitn.it,  $^b$ E-mail: pinamont@science.unitn.it\\ 

\rm\large\large

\rm\normalsize 

\rm\normalsize

 
\par 
\bigskip 

\noindent
\small 
{\bf Abstract}.  
 Local scalar QFT (in Weyl algebraic approach) is constructed on
 degenerate semi-Riemannian manifolds corresponding to
 Killing horizons in spacetime.
 Covariance properties  of the $C^*$-algebra of observables
 with respect to the conformal group $PSL(2,\bR)$ are studied.
 It is shown that, in  addition to the state
 studied by Guido, Longo, Roberts and Verch for bifurcated Killing
 horizons, which is conformally invariant and
 KMS at Hawking temperature with respect to the Killing flow
 and defines a conformal net of von Neumann algebras,
 there is a further wide class of algebraic (coherent) states
 representing spontaneous breaking of $PSL(2,\bR)$ symmetry.
 This class is labeled by functions in a suitable Hilbert space
 and their GNS representations enjoy remarkable properties.
 The states are non equivalent extremal KMS states at Hawking
 temperature with respect to the residual one-parameter subgroup of
 $PSL(2,\bR)$ associated with the Killing flow. The KMS
 property is valid for the two local sub algebras of
 observables uniquely determined by covariance and invariance
 under the residual symmetry unitarily represented.
 These algebras rely on the physical region of the manifold
 corresponding to a Killing horizon cleaned up by removing the
 unphysical points at infinity  (necessary to describe the
 whole $PSL(2,\bR)$ action).
 Each of the found states can be interpreted as a different
 thermodynamic phase, containing Bose-Einstein condensate,
 for the considered quantum field. It is finally suggested that
 the found states could describe different black holes. 
\normalsize

\section{Introduction.}
In a remarkable paper \cite{GLRV}, among other results,   
Guido, Longo, Roberts and Verch
show that, in a globally hyperbolic spacetime containing a bifurcate Killing horizon \cite{KW}, 
the local algebra of observables (realized as bounded operators associated to bounded spacetime regions 
in a suitable Hilbert space) 
may induce a local algebra of observables localized at the horizon itself with interesting properties. 
In fact, the induced local algebra  turns out to be covariant with respect  
to a unitary representation of M\"obius group of the circle $PSL(2,\bR) := SL(2,\bR)/\{\pm I\}$ defined 
in the Hilbert space of the system. The covariance property is referred to 
the geometric action of the M\"obius group of the circle
on the horizon as explaned below.
The work, on a hand uses general theorems due to
Wiesbrock  \cite{Wiesbrock:1992mg,Wiesbrock:1992rq} establishing  
the existence of $SL(2,\bR)$ representations related to modular operators
of von Neumann algebras. On the other hand it enjoys some interplay 
with several ``holographic'' ideas (including LightFront Holography) in QFT \cite{Sch,mopi02, mopi3,mopi4}.\\
The central mathematical object employed in \cite{GLRV} is a net of von Neumann algebras built upon
a certain state which is assumed to exist and satisfy the following requirement. 
Its restriction to the subnet of observables which are localized at the horizon,
must be  KMS at Hawking temperature for the Killing flow. In that case, the net of observables localized at
the future horizon $\bF$ (see fig.1) is shown to support a unitary representation of $PSL(2,\bR)$  giving rise to 
a {\em conformal net} (see for instance \cite{Buchholz, Gabbiani,GL, Carpi} and references therein).\\
It is worth noticing that the full $PSL(2,\bR)$-covariance of the observables of the conformal net is apparent  
when one extends the future Killing horizon $\bF$  by adding  points at infinity obtaining 
a manifold $\bS^1 \times \Sigma$,
$\Sigma$ being the transverse manifold at the bifurcation of horizons. 
$\bS^1$ represents nothing but the history $\bR$ of a particle of light 
living on the future  horizon compactified into a circle by means of the addition of a point at infinity. 
 The addition of  points at infinity is necessary because $PSL(2,\bR)$ acts properly as a subgroup of the diffeomorphisms of the circle $\bS^1$ and not the line $\bR$.
 In particular the action of $PSL(2,\bR)$ on $\bS^1$  includes arbitrary rotations of the 
 circle itself which shift the point at infinity in the physical region $\bR$. \\
 From a physical point of view these transformations have no meaning. So it seems that 
 the found covariance of the observables localized at the horizon under the full group $PSL(2,\bR)$
 is actually too large. The problem could be traced back to the state used to construct the von Neumann
 net of observables. \\
  In spite of this  drawback, the  results proved in \cite{GLRV} shows the existence of a nice 
  interplay of Killing horizons,  thermal states at the correct physical
 temperature, and conformal symmetry.  This  result is strongly remarkable 
in its own right.  \\
In the first part of this paper we give an explicit procedure to built up a local algebra of observables 
localized on a degenerate semi-Riemannian manifold $\bM:= \bS^1\times \Sigma$
(obtained from future or past Killing horizons  in particular) based on Weyl quantization procedure. This is done  
without referring to external (bulk) algebras and states and restriction procedures. 
We find, in fact, a conformal net of 
observables relying on a $PSL(2,\bR)$-invariant vacuum $\lambda$: 
At algebraic level there is a representation $\alpha$ 
of $PSL(2,\bR)$  made of $^*$-automorphisms of the Weyl algebra 
$\cW(\bM)$ and  there is a state $\lambda$
on $\cW(\bM)$  which is invariant under $\alpha$. In the  GNS representation of $\lambda$, $\alpha$ is
implemented  covariantly by a unitary representation $U$ of $PSL(2,\bR)$.
Moreover it is showed that $\lambda$ is KMS at Hawking temperature, with respect to the generator of conformal dilatations, in suitable 
regions $\bF_\pm$ of $\bM$ (see fig.1).  $\bF_\pm$ do not include points at infinity and are to the two 
disjoint regions in $\bF$
respectively in the past and in the future of the bifurcation surface.\\
In the second part we try to solve the problem focused above concerning the physical inappropriateness
of the full $PSL(2,\bR)$ covariance whenever $\bM$ is realized by adding (unphysical) points at infinity 
to a future 
Killing horizon $\bF$.\\
To this end, it is proven  that  it is possible to get rid of the unphysical action of $PSL(2,\bR)$ and single out
  the physical part of the horizon  at {\em quantum level,  i.e. in Hilbert space}, through a 
  sort of {\em spontaneous breaking of $PSL(2,\bR)$ symmetry}. In fact,  
   we establish  the existence of  other, unitarily inequivalent, GNS representations of $\cW(\bM)$
 based on new coherent KMS states $\lambda_\zeta$ at Hawking temperature. Here $\zeta$ denotes any functions in $L^2(\Sigma)$. 
  Those states are
 no longer invariant under the whole representation $\alpha$ and in particular they are not 
 invariant under the unphysical transformations of $PSL(2,\bR)$. However the residual
 symmetry  still is covariantly and unitarily implementable and  singles out the algebras  $\cA(\bF_+)$ and  
 $\cA(\bF_-)$ as unique invariant subalgebras. The states $\lambda_\zeta$  represent 
 different {\em thermodynamical phases} with respect to $\lambda$
(this is because the states $\lambda_\zeta$ are extremal KMS states) at Hawking temperature. 
Those states have different properties  in relation  with the appearance of a Bose-Einstein 
condensate
localized at the horizon.  Finally we  suggest that these states could, in fact, denote
different black holes. 
In this view the bosonic field $\phi$ generating the Weyl
representations could represent a noncommutative coordinate in the physical 
regions $\bF_\pm$, whereas  its mean value represents the classical 
coordinate describing the parameter of integral curves of the Killing vector 
restricted to the horizon. \\
Several comments concerning the representation of the whole
group $Dif\spa f^+(\bS^1)$ and in particular its Lie algebra in the presence of 
the transverse manifold $\Sigma$, are spread throughout the work.

\section{Scalar free QFT on degenerate semi-Riemannian manifolds.} 
\ssa{Basic definitions and notation}\label{geometry}
In this paper we deal with metric-degenerate semi-Riemannian manifolds
  of the product form  $\bS^1 \times \Sigma$,
where $\Sigma$ is a connected oriented $d$-dimensional  manifold equipped with a
positive metric. $\bS^1$ is assumed to be oriented and endowed with the null metric. $\bS^1 \times \Sigma$
itself is oriented by the orientation induced from those of $\bS^1$ and $\Sigma$.
$\bS^1 \times \Sigma$ will be called {\bf degenerate manifold} in the following
and it will be denoted by $\bM$ throughout.
A {\bf standard frame}  $\theta$ on the factor $\bS^1$ 
of $\bM$ is a positive-oriented local coordinate patch on  $\bS^1$ which maps 
  $\bS^1\setminus\{\infty\}$ bijectively
to the the segment $-\pi <\theta < \pi$, $\infty$ being a point of $\bS^1$. 
Througout $C_c^\infty(\bM; \bR)$ and $C_c^\infty(\bM; \bC)$ denote the space of 
compactly-supported real-valued,  resp. complex-valued,  smooth functions
on $\bM$ and  $\omega_\Sigma$ is the  volume form on $\Sigma$ induced by the
metric of $\Sigma$.  $C_c^\infty(\bM; \bC)$ is endowed with a natural  symplectic 
(i.e. bilinear and antisymmetric) form 
given by,  if $\psi,\psi'\in C_c^\infty(\bM; \bC)$,
\beq   
\Omega(\psi,\psi')
:=\int_{\bM} \psi' \epsilon_\psi -\psi \epsilon_{\psi'} \:\:\:\:\:\mbox{where}\:\:\:\:\: \epsilon_\psi := 
d\psi \wedge \omega_\Sigma
 \label{cf}\:.
\eeq
Concerning KMS states we adopt the definition  5.3.1 in \cite{BR2} (see also chapter V of
\cite{Haag} where the $\sigma$-weak topology used in the definition above in the case of a von Neumann
algebra is called {\em weak $\:^*$-topology} also known as {\em ultraweak topology}). \\
The symbol $\bN$  denotes the set of natural numbers $\{0,1,2,\ldots\}$, whereas 
$\bN'$ means $\bN\setminus \{0\}$.\\
 
\ssa{Bifurcate Killing horizon and Kruskal  cases} A simple example of three-dimensional 
degenerate manifold can be obtained from a submanifold 
of  Kruskal manifold.
However everything follows is valid, more generally, for any $(d+2)$-dimensional globally hyperbolic spacetime
 containing a bifurcate Killing horizon \cite{KW} if replacing $\bS^2$ with a generic $d$-dimensional spacelike
  submanifold $\Sigma$. In
  A basis of  Killing vector fields  of Kruskal spacetime
 is made of three fields: two generating the $\bS^2$ symmetry 
and $\xi$ generating time evolution in the two static open wedges where $\xi$
is timelike.
The region where $(\xi, \xi)=0$ is made of the union of two three-dimensional submanifolds, $\bP$
and $\bF$, which we call, respectively the {\bf past} and the {\bf future Killing horizon} 
of the manifold in reference to fig.1.
 $\bP\cap \bF$ is the {\em bifurcation surface}, i.e.
  a spacelike two-dimensional oriented submanifold where $\xi=0$, given by $\bS^2$ 
  equipped with the Euclidean 
  standard metric of a $2$-sphere with radius  given by Schwarzschild one $r_s$. That metric
  is induced from the spacetime metric.
 $\bF$ is isometric to the  degenerate manifold  
 $\bR \times \bS^2$. $\bR$ is made of the  orbits of the null Killing vector $\xi$ restricted to $\bF$.
 We assume that the origin of $\bR$  is  arranged to belong to the bifurcation manifold $\bS^2$.  The metric 
 induced on $\bF$ is degenerate along $\bR$ and  invariant under $\bR$-displacements.
 A degenerate manifold $\bM=\bS^1\times \Sigma$   can,  obviously,  
 be obtained from $\bF$ by adding a point at infinity $\infty$ to $\bR$ producing $\bS^1$. 
 In this case $\bM= \bS^1\times \bS^2$.  
 Orientation of $\bS^1$ is that induced by $\bR$.
 Then $\theta(V) = 2\tan^{-1}V$, with $V\in \bR$, is a 
 standard  frame on $\bS^1$. \\
Other examples of degenerate manifold arise from the event horizon of topological black-holes \cite{vanzo,Mann} where $\Sigma$
is replaced by a compact two-dimensional manifold of arbitrary nonnegative  genus.\\
\begin{center}
\epsfig{file=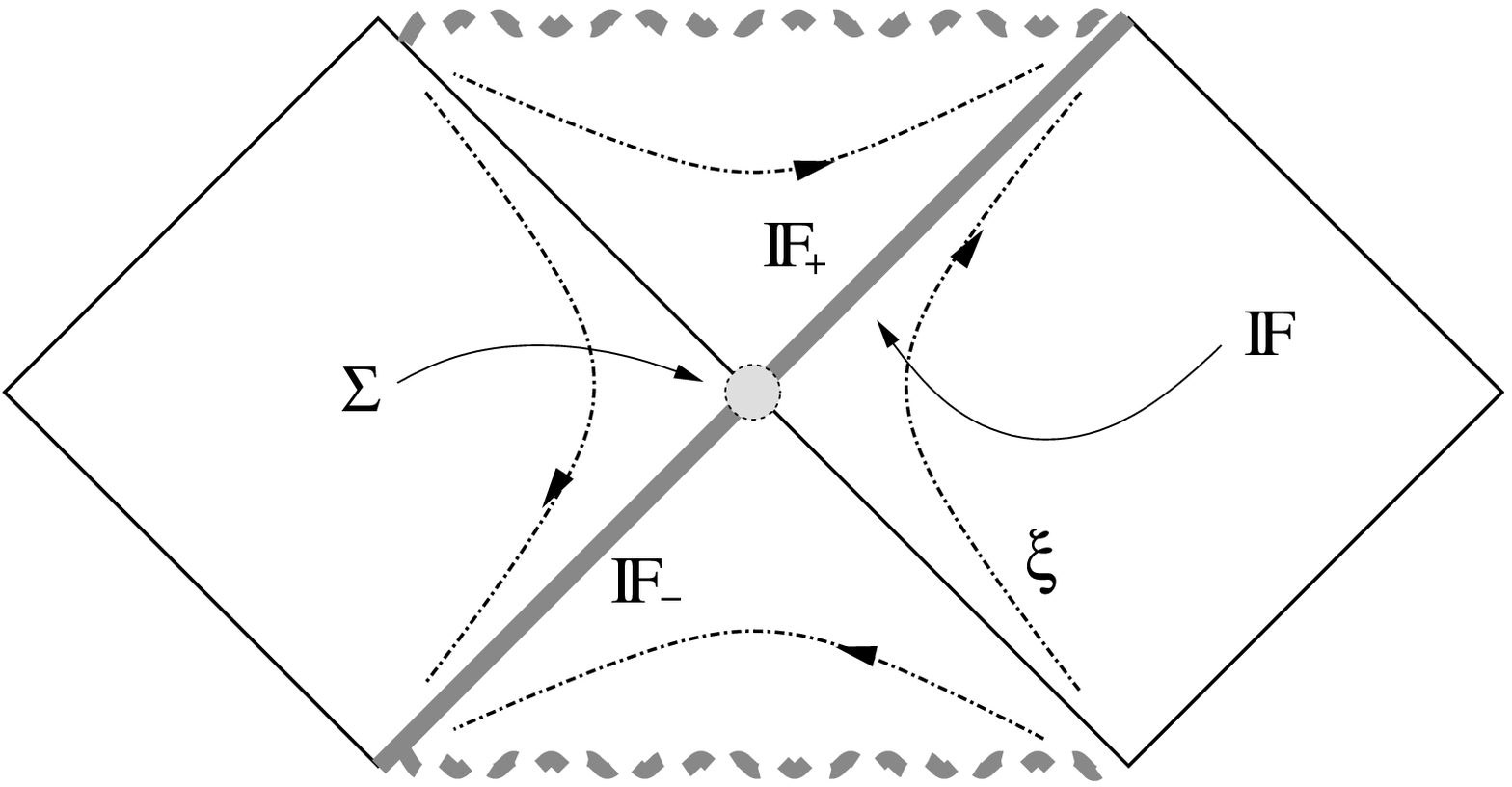, width=7cm}\\
Fig. 1. Carter-Penrose conformal diagram of Kruskal spacetime\\
\end{center}
\noindent

\ssa{Weyl/symplectic  approach} In \cite{mopi02, mopi3, mopi4} 
we have considered the limit case of a degenerate manifold $\bM=\bS^1$ where
$\bM\setminus \{\infty\}$ is as part of its boundary made of a bifurcate Killing horizon
in $2D$ Minkowski spacetime. In that case local QFT 
can be induced on $\bM$, by means of a suitable restriction 
procedure of standard
linear QFT in the bulk spacetime. This restriction actually enjoys some holographic properties
because it preserves information about bulk quantum field theory.
Here we construct QFT 
on a general degenerate manifold $\bM = \bS^1 \times \Sigma$ without referring to any restriction 
procedure.
The restriction procedure with holographic properties could be generalized to more complicated manifolds
(Kruskal manifold in particular) and this issue will be  investigated elsewhere. 
The formulation of  real scalar QFT on a degenerate manifold $\bM$ 
we present here is an adaptation of 
 the  theory of fields obeying 
 linear field equations in globally hyperbolic spacetimes \cite{BR2,KW,Wald0,Wald}\footnote{In this paper, barring 
 few differences,  we make use of conventions and notation  of  \cite{Wald}.}. 
The  starting point of QFT is  the real vector space of  {\bf  wavefunctions} 
$\cS(\bM):= C_c^\infty(\bM; \bR)/\sim$,
where   $\psi \sim\psi'$ iff $\epsilon_\psi = \epsilon_{\psi'}$.
$\Omega$ induces a symplectic form on $\cS(\bM)$ still indicated by $\Omega$ and defined by, 
if $[\psi], [\psi'] \in \cS(\bM)$,
\beq
\Omega([\psi],[\psi'])
:=\Omega(\psi,\psi') \label{zero}\:.
\eeq
\noindent{\bf Remarks}.\\
{\bf (1)} Two facts hold: (a) $\psi\sim \psi'$ iff $\frac{\partial \psi -\psi'}{\partial \rho} =0$ 
everywhere, and  (b) $\epsilon_\psi = 
\frac{\partial \psi}{\partial \rho} d\rho \wedge \omega_\Sigma$, 
$\rho$ being  any (local) coordinate on $\bS^1$.
Using (a) and (b) one proves straightforwardly that
$\Omega([\psi],[\psi'])$ is well-defined, that is it does not depend on the representatives $\psi,\psi'$
 chosen in the classes $[\psi],[\psi']$.\\
 {\bf (2)} $\Omega$ is {\em nondegenerate}  on $\cS(\bM; \bR)$, that is $\Omega([\psi],[\psi'])=0$ for all $[\psi']\in \cS(\bM)$
implies $[\psi]=0$.  The definition $\cS(\bM):= C_c^\infty(\bM; \bR)/\sim$ gets rid of  degenerateness 
of $\Omega$ on  $C_c^\infty(\bM; \bR)$ due to functions 
constant in $\bS^1$. Nondegenerateness allows the use of standard procedure to built up QFT
within the Weyl formalism as explaned below. Another possibility to remove
degenerateness is to define $\cS(\bM)$ as the
space of  $C_c^\infty(\bM; \bR)$ functions with vanishing integral with respect to some
measure $d\rho$ induced by a coordinate $\rho$ on $\bS^1$.
 Such a definition, differently to that given above, would break  invariance under orientation-preserving
 diffeomorphisms of $\bS^1$, which is a natural physical requirement due to the 
absence of a metric on  $\bS^1$. Breaking diffeomorphism invariance
will enter the theory through the choice of a reference quantum state.\\
{\bf (3)} {\em Henceforth we indicate a wavefunction $[\psi]$ by $\psi$ if the notation is not
 misunderstandable.}\\
 
\noindent ``Wavefunction'' is quite an improper term, due to the absence of any equation of motion
on $\bM$, nevertheless the ``wavefunctions'' introduced here play a r\^ole similar to that of the  smooth solutions 
of Klein-Gordon equation in a globally hyperbolic spacetime. 
As $\cS(\bM)$ is a real vector space equipped with a {\em nondegenerate}
symplectic form $\Omega$, 
there exist a complex $C^*$-algebra  (theorem 5.2.8 in \cite{BR2})
generated by elements, $W(\psi)$ with $\psi\in \cS(\bM)$
satisfying, for all $\psi, \psi' \in \cS(\bM)$,
$$\mbox{(W1)} \:\:\:\:\: \:\:\:\:\:W(-\psi) = W(\psi)^*,\:\:\:\:\: \:\:\:\:\: 
\mbox{(W2)}\:\:\:\:\: \:\:\:\:\: W(\psi)W(\psi') = e^{i\Omega(\psi,\psi')/2} W(\psi+\psi')\:.$$
That $C^*$-algebra, indicated by ${\cal W}(\bM)$,  is unique up to (isometric) $^*$-isomorphisms (theorem 5.2.8 in \cite{BR2}).
As consequences of (W1) and (W2), ${\cal W}(\bM)$ admits unit $I=W(0)$, each $W(\psi)$ is unitary and, 
from the nondegenerateness of $\Omega$, 
$W(\psi)= W(\psi')$ if and only if $\psi=\psi'$.
${\cal W}(\bM)$ is called {\bf Weyl algebra associated with} $\cS(\bM)$ and $\Omega$
whereas the $W(\psi)$ are called {\bf symplectically-smeared (abstract) Weyl
operators}.
The formal interpretation of elements $W(\psi)$ is   $W(\psi) \equiv  e^{i\Omega(\psi,\hat{\phi})}$
where $\Omega(\psi,\hat{\phi})$ are {\bf symplectically-smeared scalar fields} as we shall see shortly.\\

\ssa{Implementing locality: fields smeared with forms}  
 In a globally-hyperbolic  spacetime $X$ the {\em local} smearing is obtained employing real compactly supported functions $f$ 
 instead of solutions of field equations \cite{Wald} to smear field operator. In particular one gives a rigorous 
 meaning to:
 \beq \hat{\phi}(f) = \int_X \hat{\phi}(x) f(x)\: d\mu(x)\:,\label{hypo}\eeq
 $\mu$ being the measure induced by the metric of $X$.
 The support of the smearing function $f$ gives a suitable notion of support of the 
 associated observable  $\hat{\phi}(f)$. In this way locality can be implemented by stating that observables 
 with causally disjoint supports
 commute. In our case it is impossible to assign a unique support to a class of equivalence $[\psi]$ 
and thus implementation of locality is not very straightforward in the  symplectic 
approach. Furthermore, there is no natural measure $\mu$ on $\bM$ 
as that present in (\ref{hypo}) because $\bS^1$ is metrically degenerate. 
Both problems can be solved by using compactly supported forms instead of compactly supported functions.
 Let us indicate by $\cD(\bM)$ the space of real forms $\epsilon_\psi$ (see (\ref{cf}))
with $\psi \in C_c^\infty(\bM; \bR)$. In a globally hyperbolic spacetime  
\cite{Wald} the relation between wavefunctions   
and smooth compactly supported functions (now elements of $\cD(\bM)$)
used in (\ref{hypo}),
is implemented by the {\em causal propagator}, $E: \cD(\bM) \to \cS(\bM)$, \cite{Wald}.
It is a $\bR$-linear surjective  map which associates a smooth function with a 
 wavefunctions (supported in the causal set generated by the support of the smooth function)  
 and satisfies several properties. 
 The crucial property describing the interplay of $E$ and $\Omega$ reads, if $E(\omega,\omega') := 
 \int_{\bM} E(\omega) \omega'$,
  \beq 
  \Omega(E \omega, E\omega') = E(\omega,\omega')\:\:\:\:\:\:\mbox{for all $\omega,\omega' \in \cD(\bM)$} \:,\label{inter}
  \eeq
In our case (\ref{inter}) and
 surjectivity determine $E$ uniquely on $\bM$.\\
 
\proposizione\label{prop1}
{\em On a degenerate manifold $\bM:= \bS^1\times \Sigma$ there  is a unique  surjective $\bR$-linear  map  $E: \cD(\bM) \to \cS(\bM)$
satisfying (\ref{inter}). Moreover the following facts hold.\\
 {\bf (a)} If  $\theta$ is a standard  frame on $\bS^1$ and  $\omega \in \cD(\bM)$ is realized  as a  $2\pi$-periodic form in $\theta$
 viewed as positive-oriented coordinate $\bR$ and $s\in \Sigma$, $E$ admits the representation
\beq
(E(\omega))(\theta,s) = \left[\frac{1}{4}\int_{\theta' \in [-\pi,\pi]\: s'\in \Sigma}\left(\text{sign}(\theta')-\frac{\theta'}{\pi}\right) \delta(s,s') 
\omega(\theta-\theta',s')\right]\:. \label{tre}
\eeq
{\bf (b)} $E$ is {\em bijective} and  in particular, for $\psi \in \cS(\bM)$, $\omega \in \cD(\bM)$,  one has  
\beq E(\epsilon_\psi)= \frac{1}{2}\psi \:\:\:\:\mbox{and  $\:\:\epsilon_{E(\omega)}= \frac{1}{2}\omega$}\label{quattro}\:.\eeq
Thus $(\omega,\omega') \mapsto  E(\omega,\omega')$ is a nondegenerate symplectic form on $\cD(\bM)$.}\\

\noindent{\em Proof}. The fact that $E$ defined in (\ref{tre}) satisfies (\ref{inter})
can be proved straightforwardly by direct computation.  Direct computation shows also
the validity of (\ref{quattro}) proving injectivity and surjectivity. Any linear surjective map $E$ satisfying (\ref{inter}) fulfills also
$\Omega(\psi, E\omega') = \int_{\bM} \psi \omega'$
for every $\psi \in \cS(\bM)$ and $\omega' \in \cD(\bM)$. 
If  $E, E'$ are surjective linear maps satisfying (\ref{inter}),   one has
$ \Omega(\psi,E\omega -E'\omega)=  \int_{\bM} \psi (\omega - \omega) =0$
for every $\psi \in \cS(\bM)$.  $\Omega$ is non degenerate and thus $E\omega -E'\omega =0$ for every  $\omega \in \cD(\bM)$. Hence $E=E'$. The final statement is now obvious. $\Box$\\

\noindent We shall call the bijective map $E$  in (\ref{tre}) {\bf causal propagator},
regadless of the partial  inappropriateness of the name due to the lack of field equations. In spacetimes, existence of field equations
is responsible for the failure of the injectivity of the causal propagator.
$\delta(s,s')$ in (\ref{tre})
has an evident physical meaning  if   $(\bS^1\setminus \{\infty\})\times \Sigma$ is thought  as the future Kruskal  Killing
  horizon 
and $E$ is interpreted as the limit case of a properly defined causal propagator: As  the boundary of a causal sets $J(S)$, for $S\subset \Sigma$, 
is made of portions of the factor $\bS^1$, causal separation of sets $S,S' \subset \Sigma$ 
assigned at different ``times'' of $\bS^1\setminus \{\infty\}$, is equivalent to $S\cap S' = \emptyset$.

\noindent
As in  spacetimes,  if $\omega \in \cD(\bM)$,  
   the {\bf form-smeared (abstract) Weyl field} is defined as
   \beq
 V(\omega):= W(E \omega) \label{1form}.
\eeq
With this definition one immediately gets Weyl relations once again:  For all $\omega, \eta \in \cD(\bM)$,
$$\mbox{(V1)} \:\:\:\:\: \:\:\:\:\:V(-\omega) = V(\omega)^*,\:\:\:\:\: \:\:\:\:\: \mbox{(V2)}\:\:\:\:\: \:\:\:\:\: V(\omega)V(\eta) = e^{iE(\omega,\eta)/2}V(\omega+\eta)\:.$$
Since $E$ is injective, differently from
the extent in a spacetime, $V(\omega)=V(\omega')$ if and only if $\omega=\omega'$.
A notion of {\em locality} on $\bM$ (in a straightforward extension of  original idea  due to Sewell
\cite{Sewell}) 
can be introduced at this point by the following proposition (the proof is in the appendix).\\ 

\proposizione\label{prop2.2}
{\em $[V(\omega), V(\omega')]=0$
for $\omega,\omega' \in \cD(\bM)$ if  one  of the conditions  is fulfilled:\\
 {\bf (a)} there are two  open  disjoint segments $I,I'\subset \bS^1$ with
 $supp\: \omega \subset I \times \Sigma$ and   $supp\: \omega' \subset I' \times \Sigma$,\\
{\bf (b)}  there are two open disjoint sets $S,S'\subset \Sigma$ with
  $supp\: \omega \subset \bS^1 \times S$ and   $supp\: \omega' \subset \bS^1 \times S'$.}\\

\noindent 
The $^*$-algebra ${\cal W}(\bM)$ is {\em local} in the sense stated in the thesis of Proposition \ref{prop2.2}. Notice that $supp\: \omega \cap supp \: \omega' = \emptyset $ does not imply commutativity of $W(\omega)$ and $W(\omega')$ in general.\\

\ssa{Fock representations}
Breaking  invariance under orientation-preserving $\bS^1$-diffeomorphisms,  Fock representations of  ${\cal W}(\bM)$ can be introduced as follows
 generalizing part of the construction presented in 2.4 of \cite{GLW} and  in \cite{mopi4}. From a physical point of view, the procedure   
resembles  quantization with respect to Killing time in a static spacetime.
Fix a standard  frame $\theta$ on $\bS^1$.
Any representative $\psi$ of   $[\psi] \in \cS(\bM)$ can be expanded in Fourier series in the parameter
$\theta$, where $\bN':= \bN\setminus \{0\}$,
\beq
\psi(\theta, s)\sim \sum_{n\in \bN'} \frac{e^{-in\theta}\widetilde{\psi(s,n)}_+}{\sqrt{4\pi n}} + 
 \sum_{n\in \bN'} \frac{e^{in\theta}\overline{\widetilde{\psi(s,n)}_+}}{\sqrt{4\pi n}} \; =
\psi_+(\theta, s)+\overline{\psi_+(\theta, s)}\:.\label{ser}
\eeq
 $\psi_+$  is the  $\theta$-{\bf positive frequency part} of  $\psi$. 
 The term with $n=0$ was discarded due to the equivalence relation used defining 
$\bS(\bM)$, the remaining terms  depend on $[\psi]$ only.  $\Sigma \ni s\mapsto \widetilde{\psi(s,n)}_+$ is 
smooth, supported in a compact set of $\Sigma$ independent from $n$ and, using integration by parts,
  for any $\gamma>0$, there is $C_\gamma\geq 0$ with
  $||\widetilde{\psi(\cdot,n)}_+||_\infty \leq C_\gamma n^{-\gamma}$ for $n\in \bN'$
  so that the  series in (\ref{ser}) converges uniformly and
   $\theta$-derivative operators can be 
  interchanged with  the symbol of summation. The found estimation  and Fubini's theorem entail that
the sesquilinear form  \beq
\al \psi'_+,\psi_+\cl  :=-i\Omega(\overline{\psi_+'},\psi_+) \label{prodscalar}
\eeq
on the space of complex linear combinations of $\theta$-positive frequency parts 
 satisfies 
\beq
\al \psi'_+,\psi_+\cl  =\sum_{n=1}^{\infty} \int_{\Sigma} \overline{\widetilde{{\psi'}(s,n)}_+}\:
\widetilde{{\psi}(s,n)}_+ \:\omega_\Sigma(s)
= \int_{\Sigma} \sum_{n=1}^{\infty} \overline{\widetilde{{\psi'}(s,n)}_+}\:\widetilde{{\psi}(s,n)}_+
\:\omega_\Sigma(s). \label{say}
\eeq
Thus it is positive and defines a Hermitian scalar product.
The {\bf one-particle space}
 $\cH$ is now  defined as the completion w.r.t $\al \cdot,\cdot\cl $ of the
space of positive $\theta$-frequency parts $\psi_+$ of wavefunctions.
Due to (\ref{say}),  $\cH$ is isomorphic
to $\ell^2(\bN) \otimes L^2(\Sigma,\omega_\Sigma)$
\footnote{The construction of $\cH $ is equivalent to that performed in the approach of 
\cite{Wald} (see also \cite{KW}) using the real scalar product on $\cS(\bM)$,
$\mu(\psi,\psi'):= - Im \: \Omega(\overline{\psi_+}, \psi'_+)$ and the map $K : \cS(\bM) \ni \psi \mapsto \psi_+ \in \cH $.}.
 ${\gF_+}(\cH  )$ is the symmetrized Fock space with vacuum state $ \Psi $ and one-particle space
 ${\cal H}$.
The  {\bf field operator symplectically smeared with} $\psi \in \cS(\bM)$ and the {\bf field operator smeared with the form}   $\omega\in \cD(\bM)$ are
respectively the operators:
\beq
 \Omega(\psi,\hat{\phi}) :=  i a(\overline{\psi_+})-ia^\dagger(\psi_+) 
\:\:\:\:\:\:\mbox{and}\:\:\:\:\:\:
 \hat{\phi}(\omega) :=   \Omega(E\omega,\hat{\phi}) 
\label{rfield}
\eeq
where the operators $a^\dagger({\psi_+})$  and $a(\overline{\psi_+})$ 
($\bC$-linear in $\overline{\psi_+}$)
 respectively create and annihilate the state $\psi_+$. 
 The common invariant domain 
  of all the involved  operators 
is the dense linear manifold $F(\cH )$ spanned by the vectors with finite number of particle.
$\Omega(\psi,\hat{\phi})$ and  $\hat{\phi}(\omega)$ are essentially self-adjoint on $F(\cH )$
(they are symmetric and $F(\cH )$ is dense and made of analytic vectors)
and satisfy bosonic commutation relations (CCR): 
\beq [\Omega(\psi,\hat{\phi}),\Omega(\psi',\hat{\phi})] = -i\Omega(\psi,\psi')I 
\:\:\:\:\:\:\mbox{and}\:\:\:\:\:\:
[\hat{\phi}(\omega), \hat{\phi}(\omega')] = -iE(\omega,\omega')I \nonumber
\:.\eeq The definition 
$\hat{\phi}(\omega) :=   \Omega(E\omega,\hat{\phi})$ is here nothing but a rigorous interpretation of the formula
$\hat{\phi}(\omega) = \int_\bM \hat{\phi}(x) \omega(x)$.
    Finally the unitary operators
\beq\hat{W}(\psi) := e^{i\overline{\Omega(\psi,\hat{\phi})}}\:\:\:\:\:
\mbox{and,  equivalently, $\:\:\:\:\:\:\:\:\hat{V}(\omega) 
:= \hat{W}(E\omega) =e^{i\overline{\hat{\phi}(\omega)}}$}\label{utile}\eeq
enjoy properties (W1), (W2)  and, respectively (V1), (V2), so that they define a unitary representation
$\hat{\cal W}(\bM) $  of  ${\cal W}(\bM)$ which is also irreducible.
The  proof of these properties follows from  propositions  5.2.3 and 5.2.4 in \cite{BR2}
\footnote{There the symplectic form is $\sigma=-2\Omega$ and the field operator $\Phi(\psi_+)$ of  
prop. 5.2.3 of \cite{BR2} is $\Phi(\psi_+) = 
2^{-1/2}\Omega(J\psi,\hat{\phi})$ where $J\psi= -i\psi_+ +i\overline{\psi_+}$ if $\psi= 
\psi_+ +\overline{\psi_+}$. Notice that   $J(\cS(\bM))\subset \cS(\bM)$: 
That is false in general with other definitions of $\cS(\bM)$!}.\\
If $\Pi  : {\cal W}(\bM) \to \hat{\cal W}(\bM) $ denotes the unique 
 ($\Omega$ being nondegenerate)
 $C^*$-algebra isomorphism between those two Weyl representations,  $({\gF_+}(\cH ), \Pi ,  \Psi )$ 
coincides, up to unitary transformations,  with  the GNS triple associated with the  algebraic pure state $\lambda$
on ${\cal W}(\bM)$ uniquely defined  by the requirement (see the appendix) 
\beq\lambda (W(\psi)) := e^{-\langle \psi_+, \psi_+\rangle /2}\label{GNS}\:.\eeq

\section{Conformal nets on degenerate manifolds.}

\ssa{$Dif\spa f^+(\bS^1)$,  $PSL(2,\bR)$ and associated $^*$-automorphisms on $\bM$}
 We   recall 
here some basic notions of conformal representations on $\bS^1$. 
  Let $Vect(\bS^1)$ be the infinite-dimensional  Lie algebra of the infinite-dimensional Lie group  (see
  Milnor \cite{Milnor-leshouches})
of orientation-preserving  smooth diffeomorphisms of the circle $Dif\spa f^+(\bS^1)$.  $Vect(\bS^1)$  is the real linear space of 
smooth vector fields on $\bS^1$ whose associated one-parameter diffeomorphisms 
preserve the orientation of $\bS^1$.   $Vect^\bC(\bS^1)$
denotes   the complex Lie algebra $Vect(\bS^1) \oplus i Vect(\bS^1)$
with usual Lie brackets  $\{\cdot, \cdot\}$ and involution  $\imath : X \mapsto -\overline{X}$
for $X\in Vect^\bC(\bS^1)$, so that  $\imath(\{X,Y \}) = \{\imath(Y),\imath(X)\}$.
 $Vect(\bS^1)$ is the (real) sub-Lie-algebra of $Vect^\bC(\bS^1)$ 
 of anti-Hermitean elements with respect to  $\imath$.
 $\frak{a}$ denotes the  Lie subalgebra of $Vect^\bC(\bS^1)$
whose elements  have a finite number of Fourier component with respect a standard frame $\theta$
which is supposed to be fixed from now on.
 A  basis for $\frak{a}$  is made of fields  \begin{equation}
{\cal L}_{n} := i e^{in\theta} \:\partial_\theta,\:\:\:\:\:\:\: \mbox{with $n\in \bZ$.} \label{F_n}
\end{equation}
They enjoy the so-called {\em Hermiticity condition},
$ \imath({\cal L}_n) = {\cal L}_{-n}$ and
the well-known  {\em Virasoro commutation rules} with vanishing central charge, 
$[{\cal L}_{n},{\cal L}_{m}]= (n-m) {\cal L}_{ n+m}$. \\
We remind that $SL(2,\bR)$ and
$SU(1,1)$ are isomorphic through the map $SL(2,\bR)\ni h \mapsto g \in SU(1,1)$ where:
$$
 g:= 
\begin{pmatrix}\zeta &\overline{\eta} \\{\eta}&\overline{\zeta} \end{pmatrix},
 h = \begin{pmatrix}\a&\b\\ \c &\d \end{pmatrix}\:
 \mbox{and}\:\:\zeta:= \frac{\a+\d +i(\b-\c)}{2}\:,\:\:\: \eta:= \frac{\d-\a-i(\b+\c)}{2}\:.$$
  $Dif\spa f^+(\bS^1)$ includes the 
 {\bf M\"obius group of the circle} 
$PSL(2,\bR):= SU(1,1)/\{\pm I\}$ as a 
finite-dimensional subgroup: Thinking $\bS^1$ as the unit complex circle paremetrized by $\theta$,
 an element $g \in PSL(2,\bR)$ is injectively associated with the
 diffeomorphism $g\in Dif\spa f^+(\bS)$ 
\beq
g : e^{i \theta}
  \mapsto
 \frac{\zeta e^{i\theta} +\overline{\eta}}{\eta e^{i\theta} +\overline{\zeta}}, \:\:\:\:\mbox{with}\:\:\:\: 
\theta \in [-\pi,\pi]\:.\label{moebius}
 \eeq 
The corresponding inclusion of Lie algebras is illustrated by the fact that  the three $\imath$-anti-Hermitean linearly-independent
 elements of  $\frak{a}$ 
 \beq\label{vecKSD}
{\cal K}:=i{\cal L}_0 = -\partial_\theta\:, \:\:\:\:{\cal S}:= i\frac{{\cal L}_1 + {\cal L}_{-1}}{2} = -\cos \theta \partial_\theta
\:,\:\:\:\:{\cal D}:= i\frac{{\cal L}_{1} -  {\cal L}_{-1}}{2} = -\sin \theta  \partial_\theta\label{vectors}
\eeq
enjoy  the  commutation rules  of the elements $k,s,d$ of  the basis  of the Lie algebra $sl(2,\bR)$
with
 \begin{eqnarray}
 k =  \frac{1}{2}\left[
\begin{array}{cc}
  0 &1\\
  -1 & 0 
\end{array}
\right]\:,\label{si1}
\:\: s = \frac{1}{2}\left[
\begin{array}{cc}
 0 & 1\\
 1 & 0 
\end{array}
\right]\:,\label{si2}
\:\: d = \frac{1}{2}\left[
\begin{array}{cc}
1 & 0 \\
0 & -1
\end{array}
\right]\:.\label{si3}
 \end{eqnarray}
In particular $k$ is the generator of the {\bf subgroup of rotations} $SO(2)/\pm I\subset PSL(2,\bR)$ given by dispacements
in $\theta$. 
$Dif\spa f^+(\bS^1)$ acts naturally as a group of 
{\em isometries} on  the semi-Riemannian manifold $\bM= \bS^1 \times \Sigma$. 
If $g\in Dif\spa f^+(\bS^1)$, we shall use the same symbol to indicate the associated diffeomorphism of
$\bM$.\\

\ssa{Invariance with respect to  $PSL(2,\bR)$} 
 From now on we use the following notation. If $g \in Dif\spa f^+(\bS^1)$ and $\psi \in C^\infty_c(\bM;\bC)$, $\psi^{(g)} := \psi\circ g$.  
If $[\psi] \in \cS(\bM)$, the element $[\psi]^{(g)}  := [\psi^{(g)}]$
is well defined and it will be indicated by $\psi^{(g)}$ simply if the meaning is clear from the context. The usual pull-back 
action on forms $\omega \in \cD(\bM)$ will be denoted similarly: $\omega^{(g)}:= g^*\omega$. Notice that $g^*$ leaves $\cD(\bM)$ fixed:
Using  (\ref{quattro}), it results that  if $\psi= E\omega$ with $\omega \in \cD(\bM)$
then $\omega^{(g)}= 2\epsilon_{\psi^{(g)}} \in \cD(\bM)$.\\ 
$\Omega$ and $E$ are {\em invariant} under $Dif\spa
f^+(\bS^1)$. That is, 
 for all $\psi,\phi\in  C^\infty_c(\bM;\bC)$,  $g \in Dif\spa f^+(\bS^1)$
and $\omega,\eta\in \cD(\bM)$, 
\beq
\Omega(\psi,\phi)= \Omega(\psi^{(g)},\phi^{(g)}) \:\:\:\:\:\mbox{and}\:\:\:\:\:
E(\omega,\eta) =  E(\omega^{(g)},\eta^{(g)})\:.\label{added}\eeq
Therefore, as a consequence of general results ((4) in theorem 5.2.8 of \cite{BR2}), 
$Dif\spa f^+(\bM)$ admits  a representation $\alpha: g\mapsto \alpha_g$ made of  
$^*$-automorphisms of the algebra 
${\cal W}(\bM)$
induced by 
\beq
\alpha_g(V(\omega)) := V(\omega^{(g^{-1})})\:.\label{inv3}\eeq
In the following we employ only {\em the restriction} of the representation $\alpha$ to the
M\"obius group of the circle $PSL(2,\bR)\ni  g \mapsto \alpha_g$
in terms of  
$^*$-automorphisms of  ${\cal W}(\bM)$.\\
The definition of the state $\lambda$ (\ref{GNS}) is not $Dif\spa f^+(\bS^1)$
invariant since it relies upon the choice of a preferred standard frame $\theta$. Let us show that actually a different 
standard frame $\theta'$ produces the same 
$\lambda$ provided the coordinate transformation $\theta'=\theta'(\theta)$ 
belongs to 
$PSL(2,\bR)$. \\

\teorema\label{0} {\em Let  $\theta$ be a standard  frame on $\bS^1$ of  
$\bM:= \bS^1\times \Sigma$, consider the state on $\cW(\bM)$, $\lambda$ (\ref{GNS})
and the representation $\alpha$ of $PSL(2,\bR)$ defined above. The following hold.\\
{\bf(a)} $\lambda$ is invariant under $\alpha$, that is 
$\lambda(\alpha_g(w)) = \lambda(w)$ for all $g\in PSL(2,\bR)$ and $w\in \cW(\bM)$.\\
{\bf (b)} If $\theta'$ is another standard frame $\bS^1$ such that the coordinate 
transformation $\theta'=\theta'(\theta)$ belongs to $PSL(2,\bR)$, then $\lambda'=\lambda$
where $\lambda'$ is the analog of $\lambda$ referred to $\theta'$}.\\
 
\noindent  The proof arises from (\ref{GNS}) using  the invariance of $\Omega$ under $Dif\spa f^+(\bS^1)$ 
and the following lemma. \\

 \noindent \lemma\label{lemma1} {\em Let  $\theta$ be a standard frame on $\bS^1$ of  
$\bM:= \bS^1\times \Sigma$. The action of $PSL(2,\bR)\subset Dif\spa f^+(\bS^1)$ preserves positive frequency parts.
That is, if $g\in PSL(2,\bR)$, $\psi \in \cS(\bM)$,
$\omega \in \cD(\bM)$,
\beq
(\psi^{(g)})_+ = (\psi_+ \circ g) \:\:\:\:\mbox{and}\:\:\:\: (\omega^{(g)})_+ = g^* \omega_+ \label{oldresult}\:,
\eeq
where $\omega_+:= \epsilon_{\phi_+}$  is called the {\bf $\theta$-positive-frequency part} of 
any form  $\omega := \epsilon_{\phi}$ in  $\cD(\bM)$.}\\

\noindent{\em Proof}. From {\em Remark} on p. 271 of \cite{mopi4} one finds that
$(\psi^{(g)})_+ = (\psi_+ \circ g)$ for all $g\in PSL(2,\bR)$ and $\psi\in \cS(\bM)$. 
The result straightforwardly  extends  to  $\omega\in \cD(\bM)$  using the definition of 
$\cD(\bM)$. $\Box$\\

\noindent We stress that (\ref{oldresult}) does {\em not} hold for generic diffeomorphisms  $g\in Dif\spa
f^+(\bS^1)$.\\
 
\ssa{Virasoro representations and Conformal nets}
Let us investigate on the existence of operator representations of Virasoro algebra and the real sub algebra
$sl(2,\bR)$ in the Fock space $\gF_+({\cal H})$ introduced above focusing, in particular, on the 
relationship with the algebra $\hat{\cal W}(\bM)$.
Fix a standard  frame $\theta$ on $\bS^1$ and  build up the associated Fock
 space and the Weyl representation.
It is possible to introduce in ${\gF_+}(\cH )$ a new class of operators
which generalizes chiral currents straightforwardly. 
If $\bN':= \{1,2,3,\ldots\}$ and  $\{u_j\}_{j\in \bN'}$ is a Hilbert basis of $L^2(\Sigma,\omega_\Sigma)$ 
the vectors 
$$Z_{jn}(\theta, s) := \frac{u_j(s) e^{-in\theta}}{\sqrt{4\pi n}}$$
 define a Hilbert basis of  the one-particle space $\cH $. We can always reduce to the case of  {\em real} 
 vectors $u_j$ and  we assume that\footnote{$L^2(\Sigma,\omega_\Sigma)$ is separable since the Borel measure induced by
  $\omega_\Sigma$ is $\sigma$-finite 
and
the Borel $\sigma$-algebra of $\Sigma$ is countably  generated (the topology of $\Sigma$  being second countable by definition of manifold). If $\{u_j\}$ is a Hilbert basis $\{\overline{u_j}\}$ is such. Orthonormalization procedure of a maximal set of linearly independent  generators in the set of all $u_j+\overline{u_j}$,  $i(u_j-\overline{u_j})$ yields   a 
real Hilbert basis.} henceforth. 
The functions $\cD(\bM)\ni \omega \mapsto a(E\omega)$
and $\cD(\bM)\ni \omega \mapsto a^\dagger(E\omega)$, where the operators 
  work on the domain $F(\cH)$, can be proved to be distributions using the strong-operator topology (to show it essentially use (1) in prop. 5.2.3 in \cite{BR2})
and the usual test-function topology on $\cD(\bM)$ induced by families of  seminorms
referred to  derivatives (of any order) in coordinates of components of forms  $\omega$ (see 2.8 in \cite{Friedlander}). $\cD(\bM)\ni \omega \mapsto \hat\phi(\omega)$  admits the  distributional kernel 
\beq
\hat{\phi}(\theta,s) = \frac{1}{i\sqrt{4\pi}}\sum_{(n,j)\in \bZ\times \bN'}  \frac{u_j(s) e^{-in\theta}}{n} J^{(j)}_n \label{phicurrents}\:,
\eeq
where the {\bf (generalized) chiral currents}  $J^{(j)}_n: F(\cH ) \to F(\cH )$
are defined as follows $$J^{(j)}_0=0\:,\:\:\:\:
J^{(j)}_n =i\sqrt{n}a(\overline{Z_{jn}})\:\:\:\: \mbox{if $n>0$ and} \:\:\:\:J^{(j)}_{n} =-i\sqrt{-n}a^{\dagger}(Z_{j,-n})\:\:\:\: \mbox{if $n<0$}\:.$$
They satisfy on $F(\cH )$  both the Hermiticity condition $J^{(j)\dagger}_{n}\spa\rest_{F(\cH )} = J^{(j)}_{-n}$
and  the oscillator  commutation relations $[J^{(j)}_n,J^{(i)}_m] = n \delta^{ij}\delta_{n,-m}I$.
Introducing the usual normal order  prescription $:\cdot\cdot:$ ``operators $J^{(j)}_p$ with negative index $p$ must precede those with positive index $p$'',  one can try to define the linearly-independent  operators,
with $c\in \bN'\cup \{\infty\}$
\beq
L^{(c)}_k :=\frac{1}{2} \sum_{n\in \bZ, j\leq c} :\spa J^{(j)}_{n}J^{(j)}_{k-n}\spa:\:,\:\:\:\:\:\:\: L_k:= 
L^{(\infty)}_k \label{Lk}
\eeq
on some  domain in $\gF_+(\cH )$. We shall denote the complex infinite-dimensional algebra 
spanned by $L^{(c)}_k$ by $\hat{d}_c$.
One can formally show that  $L_k$  have two  equivalent geometric expressions
\beq 
 L_k &=& \frac{1}{2i}:\spa\Omega(\hat{\phi}, {\cal L}_k(\hat{\phi}))\spa: \label{useful}\\
 L_k &=& \int_{\bM} \spa\spa \spa: \spa \partial_\theta\hat{\phi}\partial_\theta \hat{\phi}\spa: (\theta,s) 
 e^{ik\theta}\:\: d\theta \wedge\omega_\Sigma \label{useful2} \:,
 \eeq
  ${\cal L}_k(\hat{\phi})$ is the ``scalar field'' obtained by the action of the differential operator
 ${\cal L}_k$ (naturally extended from $\bS^1$ to the product 
 $\bM= \bS^1\times \Sigma$) on the ``scalar field'' ${\phi}$. 
 The same formulae hold if replacing $L^{(c)}_k$ for $L_k$ and replacing $\hat{\phi}$ with
 $\hat{\phi}^{(c)}$ given by the right-hand side of (\ref{phicurrents}) with the sum over $j$
 restricted to the set $\{1,2,\cdots, c\}$. 
 If $c$ is finite the following proposition can be proved
by direct inspection.\\

\proposizione\label{prop3} {\em Fix a standard frame $\theta$ on $\bS^1$ 
of $\bM= \bS^1\times \Sigma$, take  $c\in \bN'$ and consider the real 
vector space $\hat{\frak{a}}_c$ generated by
 the operators $L^{(c)}_n$ in  (\ref{Lk}) equipped with the commutator $\:[\cdot,\cdot]$ 
 and the involution $\hat{\frak{a}}_c \ni a \mapsto a^\dagger\spa\rest_{F(\cal H)})$. The following holds. \\ 
 {\bf (a)} The elements of  $\hat{\frak{a}}_c$ are  well defined on 
$F({\cal H})$ which is a dense invariant space of common  analytic vectors.\\
{\bf (b)}  $(\hat{\frak{a}}_c, \:[\cdot,\cdot], \:\cdot^\dagger\spa\rest_{F(\cal H)})$ is  a  
 central representation, with  central charge $c$, of the algebra 
$(\frak{a}, \{\cdot,\cdot\},\imath)$ 
(that is a unitarizable Virasoro representation) since the following relations hold:
\begin{eqnarray}
 {L^{(c)}}_{-n} &=& {L^{(c)}_n}^\dagger\spa\rest_{F({\cal H})}\:,\\ 
\label{her}\mbox{$[$}L^{(c)}_{n},L^{(c)}_{m} \mbox{$]$} &=& (n-m) {L^{(c)}}_{n+m} +  \frac{(n^3-n)c}{12}\delta_{n+m,0}I\:. \label{virc=1}
\end{eqnarray} 
{\bf (c)} The representation is  {\em positive energy}, i.e. the generators of rotations $L_0^{(j)}$ is non-negative.\\
{\bf (d)} Each operator  $L^{(c)}_n$  does not depend on the choice for the real base $\{u_j\}_{j\leq c}$ (but depends  on the finite dimensional subspace spanned by those vectors).}\\

\noindent Notice that the found  Virasoro representations are strongly reducible \cite{Kac}. 
Once they are decomposed into unitarizable  irreducible highest-weight representations \cite{Kac}, they can be exponentiated  
(\cite{Gowa1985,Lare1999,Carpi}) obtaining unitary strongly continuous 
representations of $Dif\spa f^+(\bS^1)$.\\
In general there is 
no physical reason to single out  a Hilbert basis $\{u_j\}$ or equivalently a sequence
$\ldots {\cal H}_k \subset {\cal H}_{k+1} \ldots$
 of finite dimensional subspace  of $L^2(\Sigma,\omega_\Sigma)$.   In the presence of particular 
 symmetries   for $\Sigma$ a class of finite dimensional subspaces can be picked out  
 referring to the invariant subspaces with respect to a unitary representation on $L^2(\Sigma,\omega_\Sigma)$ of  the symmetry group.       For instance, think to $\Sigma = \bS^2$,
 in that case one may decompose $\psi \in L^2(\bS^2)$ using (real and imaginary parts of) spherical harmonics $Y^l_m$. 
Hence a suitable class of finite dimensional subspaces are those  with fixed angular momentum
$l=0,1,2,\ldots$. The sphere $\bS^2$ is reconstructed as a sequence of {\em fuzzy spheres} (\cite{Madore}) 
with greater and greater angular momentum $l$. The associated  Virasoro representations have 
central charges $c_l = 2l+1$.   

In the absence of symmetries only the case $c= \infty$ seems to be physically interesting.
Let us turn attention on this case.  Serious problems arise when trying to give a rigorous meaning to all
the operators $L_n$.
First of all (\ref{virc=1}) becomes meaningless due to $c=\infty$ in the right-hand side. Furthermore, by
direct inspection one finds  that,  
if $n< -1$, the domain of $L_n$ cannot include any vector of $F(\cH )$ due to an evident divergence  (this
drawback would arise also for $|n|=1$ if $J_0^{(j)} = 0$ were false).  
However,  by direct inspection, one finds that $L_n$ with $n\geq -1$ are well defined on 
$F(\cH )$ which is, in fact a common  invariant dense domain made of analytic vectors,
moreover $L_n\Psi=0$.
The central charge does not appear considering commutators of those operators.
The complex space (finitely) spanned by those vectors is closed with respect to 
the commutator but, unfortunately, it is {\em not} with respect to the Hermitean conjugation
so that they cannot represent a Lie algebra of observables.
However, restricting to the case $|n|\leq 1$ everything  goes right and one gets a Lie algebra
closed with respect to the Hermitean conjugation.
Anti-Hermitean linearly-independent  operators generating that Lie algebra are
 \beq\label{genKSD}
iK:=iL_0\:, \:\:\:\: iS:=i\frac{L_1 + L_{-1}}{2}\:,\:\:\:\: iD:= \frac{L_{1} -  L_{-1}}{2}\label{OP}\:.
\eeq
They enjoy  the  commutation rules  of the elements $k,s,d$ of  the basis of the Lie algebra $sl(2,\bR)$
 (\ref{si3}).
As a consequence a representation $R  : sl(2,\bR) \to {\cal L}(F(\cH ))$ can be realized by assuming
$iK = R (k), iS= R (s), iD= R (d)$  and
 $R : \alpha k+ \beta s +\gamma d \mapsto \alpha iK+ \beta iS+\gamma iD$
  for all $\alpha,\beta,\gamma \in \bR$.
One expects that this representation is associated, via exponentiation, with a strongly continuous (projective) unitary 
representation of the universal covering of  $SL(2,\bR)$, $\widetilde{SL(2,\bR)}$. Let us prove that such a
representation does exists and enjoys remarkable properties. \\

\teorema\label{1}
{\em Fix a standard  frame $\theta$ on $\bS^1$ of 
$\bM= \bS^1\times \Sigma$ and construct  the GNS (Fock) realization of $\cW(\bM)$ associated with the state $\lambda$
in (\ref{GNS}) and the representation  $R$. It turns out that the Hermitean operators  
$iR (x)$,  with $x\in sl(2,\bR)$, are essentially selfadjoint on $F(\cH )$ and
there is a unique strongly-continuous representation  $PSL(2,\bR) \ni g\mapsto U (g) : \gF_+(\cH ) \to  \gF_+(\cH )$
with
\beq U (\exp(tx)) = e^{t\overline{R (x)}}\:,\:\:\:\:\:\mbox{for all $x\in sl(2,\bR)$ and $t\in \bR$.}\label{Ne}\eeq
The following further facts hold.\\
{\bf (a)} $U $ is a {\em positive-energy} representation of $PSL(2,\bR)$ -- 
 that is the self-adjoint generator $\overline{K}$ of the subgroup of rotations,   
 has nonnegative spectrum --
and furthermore $\sigma(\overline{K}) = \{0,1,2,\ldots\}$.\\
{\bf (b)}  $U $ and its generators
do not depend on the choice of the basis $\{u_j\}_{j\in \bZ}\subset L^2(\Sigma, \omega_\Sigma)$.\\
In particular,  $U $ is  the tensorialization of $U \spa\rest_{{\cal H} }$.
Referring to  the factorization  of the one-particle space
${\cal H}  = \ell^2(\bC)\otimes L^2(\Sigma,\omega_\Sigma)$,
 it holds
$U \spa \rest_{{\cal H} } =V \otimes I\:, $
where
$V$ is the  restriction to the one-particle space 
of the representation $U$ in the simplest case  $\bM= \bS^1$. \\
{\bf (c)}
Each subspace of $\gF_+(\cH )$  with finite number of particles is invariant under 
$U $.\\
{\bf (d)} The GNS representative of $\lambda$, $\Psi$, is invariant under $U$ and  it is the only  unit vector of $\gF_+(\cH )$  invariant  under $\{e^{it\overline{D}}\}_{t\in \bR}$ 
 up to phases.}\\

\noindent The proof of the theorem is given in the appendix.
The following  further  theorem states that   $\hat{\cal W}(\bM) $
transform covariantly under this representation with respect to the action of the diffeomorphisms of 
$PSL(2,\bR)\subset Dif\spa f^+(\bS^1)$ seen in 3.2.  \\

\teorema\label{2}
{\em With  hypotheses and notation of theorem \ref{1}, the following holds.\\
{\bf (a)}  $U$ is  $PSL(2,\bR)$ covariant. In other words it  implements unitarily the 
representation $\alpha$ of  $PSL(2,\bR)$ defined in 3.2: For all
$g\in PSL(2,\bR)$,
\beq
U(g) \:w\: U(g)^\dagger
= \alpha_g(w)\:,\:\:\:\: \mbox{for all $w\in \hat{\cal W} (\bM)$} \label{a}\:.
\eeq
{\bf (b)} The one-parameter group of  $\null^*$-automorphisms associated with 
 the one-parameter group of  diffeomorphisms respectively generated by vector fields ${\cal K}, {\cal S}, {\cal D}$ correspond, trough (\ref{a}), to the  one-parameter  unitary 
 subgroups  of $U $ respectively generated by $iK,iS,iD$\footnote{Sign conventions should be clear, anyway to fix them notice that formally 
 $[iK, \hat{\phi}(\theta,s)] = -\partial_{\theta}\hat{\phi}(\theta,s)$.}.}\\

\noindent The proof of the theorem is given in the appendix. 
Theorems \ref{1} and \ref{2} has a remarkable consequence concerning the existence of a so-called {\em conformal net on
$\bS^1$} associated with the the algebra $\hat{\cW} (\bM)$. This fact has a wide
spectrum of relevant consequences in physics and in mathematics, see for instance \cite{Buchholz, Gabbiani,
GL, Carpi} and references therein.
We remind the reader that any weakly-closed $^*$-subalgebra of the unital $C^*$-algebra of all bounded operators 
on a Hilbert space is called {\em von Neuman algebra} if it contains the unit operator.  
For several theoretical reasons
(see \cite{Haag}) the largest set of bounded observables of a quantum system represented in a Hilbert space
may be assumed to be made of the  self-adjoint elements of a suitable von Neumann algebra. 
If $X$ is a $*$-algebra of bounded operators over a Hilbert space,  $X'$ denotes the algebra
of the bounded operators which commute with each  element of $X$ and it results that  \cite{Haag} 
$X$ is a von Neumann algebra if and only if $X=(X')'$. In any cases,
$X'':= (X')'$ is the minimal von Neumann algebra which contains $X$.
It is called the {\em von Neumann algebra generated by $X$}.\\

\definizione\label{def1}
{\em Let  ${\cal I}$ be  the set of non empty, nondense, open intervals of $\bS^1$. Assume that $\bS^1$ is equipped with a standard
coordinate frame $\theta$.
A {\bf conformal net on $\bS^1$} is any
triple $({\cal A},\Psi,U)$ where  ${\cal A}$ is any family $\{{\cal A}(I)\:\:|\:\: I\in {\cal I}\}$ of von Neumann algebras  on an
 infinite-dimensional separable complex Hilbert space ${\cal H}_{\cal A}$, and the following properties hold.\\
(C1) {\bf Isotony}.  ${\cal A}(I)\subset {\cal A}(J)$, if $I\subset J$ with $I,J\in {\cal I}$.\\
(C2) {\bf Locality}.   ${\cal A}(I)\subset {\cal A}(J)'$, if $I\cap J = 
\emptyset$ with $I,J\in {\cal I}$. \\
(C3) {\bf  M\"obius covariance}.  $U(g){\cal A}(I)U(g)^\dagger = {\cal A}(gI)$, $I\in {\cal I}$, $g\in PSL(2,\bR)$, where $U$ is a strongly continuous unitary representation  of 
$PSL(2,\bR)$ in ${\cal H}_\cA$ and  $g$ denotes the 
M\"obius transformation (\ref{moebius}) associated with 
$\theta$.\\
(C4)  {\bf  Positivity of the energy}. The representation $U$ is a positive-energy representation.\\
(C5)  {\bf  $U$-invariance and uniqueness of the vacuum}. $\Psi\in {\cal H}_{\cal A}$ is the  unique (up to phases) unit 
vector invariant under $U$.\\
(C6)  {\bf  Cyclicity of the vacuum}. $ \Psi$ is cyclic for the algebra ${\cal A}(\bS^1)
:= \bigvee_{I\in {\cal I}} {\cal A}(I)$.}\\

\noindent We have the following theorem.\\

\teorema\label{3}
{\em Fix a standard frame $\theta$ on $\bS^1$ of  $\bM= \bS^1\times \Sigma$ 
and define the associated Weyl algebra $\hat{\cal W}(\bM)$ in the  Fock space 
$\gF_+({\cal H})$ with vacuum state $ \Psi$ and the representation of $PSL(2,\bR)$, $U$
of Theorems \ref{1} and \ref{2}. With those hypotheses
 the family \beq
{\cal A} = 
\{{\cal A}(I)\:\:|\:\: I\in {\cal I}\} \:,\:\:\:\: \mbox{with}\:\:\:\:{\cal A}(I)=\{\hat{V}(\omega)\:\:|\:\: 
supp\: \omega \subset I\times \Sigma\}'' \label{A}\:,
\eeq
together with $\Psi$ and  $U$ form  a conformal net on $\bS^1$ 
such that  $\hat{\cW}(\bM)\subset 
{\cal A}(\bS^1)$.}\\

\noindent{\em Proof}.  (C1),  (C2) and (C3) are straightforward consequences of the definition (\ref{A}) using the fact that 
(von Neumann's density theorem)
${\cal A}(K)$ is the closure
with respect the strong operator topology 
of  the $^*$-algebra generated by the elements  in $\{\hat{W}(\omega)\:\:|\:\: supp\: \omega \subset K\times \Sigma\}$, 
employing  proposition \ref{prop2.2} concerning (C2) 
and theorem  \ref{2}
concerning (C3). 
(C4) and  (C5) are part of 
 theorem \ref{1}. (C6) is a consequence of the fact that $\Psi$ is cyclic with respect to $\hat{\cal W}(\bM)$ 
(see the appendix)
and $\hat{\cal W}(\bM) \subset {\cal A}(\bS^1)$. This inclusion is a consequence of the fact that, if $I,J\in {\cal I}$
and $\bS^1= I\cup J$, then, due to  (W2),  each element of  $\hat{\cal W}(\bM)$ has the form $c\hat{W}(\omega)\hat{W}(\omega')$ where
$supp\: \omega \subset I\times \Sigma$, $supp\: \omega' \subset J\times \Sigma$ and $|c|=1$, so that  
$\hat{\cal W}(\bM) \subset {\cal A}(I)\vee {\cal A}(J)\subset {\cal A}(\bS^1)$. $\Box$\\

\noindent
 
 \noindent {\bf Remarks}.\\
{\bf (1)} Our construction of a conformal net for, in particular, a bifurcate Killing  horizon in a globally hyperbolic
spacetime, is explicit
in giving  the effective form of of the unitary representation of $PSL(2,\bR)$ 
and the relationship with the whole
Virasoro algebra.
It  does not require any assumption on the existence of any algebra of observables in the spacetime where 
$(\bS^1 \setminus \{\infty\})\times \Sigma$ can be viewed to be  embedded, or any KMS state on that algebra.
 A different approach was presented
in \cite{GLRV} where it is shown that, in a globally hyperbolic spacetime containing a bifurcate Killing horizon, 
a conformal net can be obtained by restriction to the  horizon 
of a local algebra in the spacetime realized using a GNS representation with cyclic vector which satisfies the KMS 
condition with respect to the Killing time flow. The unitary representation of $PSL(2,\bR)$ was obtained there
making use of relevant results by Weisbrock et al \cite{Wiesbrock:1992rq,Wiesbrock:1992mg,GLW} on the interplay of modular theory and 
conformal theory. It seems plausible that our construction can be recovered also using the approach of \cite{GLRV}
defining  a bulk algebra of observables and a KMS state appropriately. This topic  will be investigated elsewhere.\\
 {\bf (2)}  Conformal nets enjoy relevant properties
\cite{Buchholz, Gabbiani, GL, Carpi}: \\
 {\bf Reeh-Schlieder property}. $\Psi$ is cyclic and separating for every $\cA(I)$.\\
  {\bf Bisognano-Wichmann property}. 
 The modular operator $\Delta_I$ associated with every $\cA(I)$ satisfies  
 $\Delta_I^{it}= U(\exp(2\pi \cD_I))$ for every $t\in \bR$, $\{\exp(t D_I)\}_{t\in \bR} \subset PSL(2,\bR)$  being the  one-parameter subgroup which leaves $I$
 invariant (with $\cD_I$ defined as in remark (2) after theorem \ref{4} below)
so that  $\Psi$ is a KMS state for $\cA(I)$ at inverse temperature $2\pi$ w.r.to $-\cD_I$ for $\cA((0,\pi))$.
\\ {\bf Haag duality}. $\cA(I)'= \cA(Int(\bS^1\setminus I))$ for every $\cA(I)$.\\
 {\bf Irreducibility}. $\cA(\bS^1)$ includes all of bounded operators on $\cH_\cA$.\\
 {\bf Factoriality}. Each $\cA(I)$ is a type $I\spa I\spa I_1$ factor.\\
 {\bf Additivity}. For every $\cA(I)$, it holds  $\cA(I)\subset \vee_{J\in S}\cA(J)$ if  $\cup_{J\in S}J \supset I$.\\  
{\bf (3)} With obvious changes,  theorems \ref{1}, \ref{2}, \ref{3} are still valid if one considers operators $L^{(c)}_{n}$ with $c<\infty$,  $|n|\leq 1$
 and the real  basis $u_j$ is made of smooth functions 
  with $j\leq c$. \\

\section{Spontaneous breaking of $SL(2,\bR)$ symmetry  and thermal states.} 

\ssa{Back to Physics} Consider the degenerate manifold 
$\bM= \bS^1\times \bS^2$ obtained by the future Killing horizon $\bF\equiv \bR\times \bS^2$
of the Kruskal manifold as discussed in section \ref{geometry}. (However what we go to say 
can be generalized to globally hyperbolic spacetimes with a bifurcate Killing horizon.)
In particular the orientation of $\bS^1 = \bR\cup \{\infty\}$ is that induced on $\bR$
from time-orientation of the spacetime.
Let $\theta$ be a standard frame on 
$\bS^1$ such that that, with $\cD$ given  in (\ref{vectors}),  
 \beq\xi\rest_{\bF} = -\kappa \cD 
 \:, \label{relation}
 \eeq
 $\xi$ being the global Killing field defining Schwarzschild time in both static wedges 
 and $\kappa$ being the {\em surface gravity} which is constant on 
 the Killing horizon $\bF$,  $\kappa = (4GM)^{-1}$, $M$  being 
 the mass of the black hole \cite{Wald0,Wald}.
 {\em Eq. (\ref{relation}) does not fix a standard frame 
  uniquely. 
  However the remaing freedom does not affect the construction we go 
 to present as consequence of theorem \ref{6} below.}  \\
 The requirement (\ref{relation}) implies that  the {\em adimensional} parameter $v\in \bR$ of the integral curves of $-\cD$ 
 on $\bF$
 coincides, up the factor $\kappa^{-1}$ and the choice for the origin,
 with the usual light-coordinate\footnote{This fact is evident using well-known global  Kruskal
 null coordinates $U,V$ \cite{Wald0}.}:
  $v= \kappa (t + r_*)$. There $r_*$ is the usual {\em Regge-Wheeler tortoise coordinate}
  and $t$ the Schwarzschild time, that is the parameter of the integral curves of $\xi$ 
  in any Schwarzschild wedge. In our picture the point $\infty$ of $\bS^1= \bR \cup \{\infty\}$ 
 corresponds to $\theta=\pi$ 
whereas $\theta=0$ corresponds to the bifurcation surface of $\bF$ (see section \ref{geometry}).\\
   Let us illustrate the physical consequences of the choice (\ref{relation}) for 
bosonic QFT  built up on the future Horizon  together with a M\"obius-covariant representation 
of $PSL(2,\bR)$ everything associated
 with the preferred choice for the coordinate $\theta$ on $\bS^1$.\\
 A celebrated  result by Kay and Wald  \cite{KW} states  that: 
 Any globally-defined  quasifree state on a globally hyperbolic spacetime with a bifurcate Killing horizon  
  (Kruskal manifold in particular) which is  invariant under $\xi$
and satisfies some further requirement ({\em Hadamard condition} inposed on the two-point function
of the quasifree state in particular) \cite{KW}  must be unique and KMS  with respect 
 to $\xi$  with the {\em Hawking inverse temperature} $\beta_H = 2\pi/\kappa$.
 From a physical point of view,
 one expects that the system of the field defined on the horizon be in thermal equilibrium with the state in the bulk.
 More precisely, since $\partial_{\kappa t}$ reduces to 
 $-\cD$ on the future horizon due to (\ref{relation}), 
  one might assume that the natural state on the Killing horizon 
is  a KMS state with respect to $-\cD$ at the inverse temperature $2\pi$:
 That coincides with Hawking inverse temperature referred to the adimensional
 ``time'' $v$ on $\bF$. 
  A first-glance candidate for such a state is just
   the restriction to $\lambda$ to the algebra of observables supported in the future Killing horizon 
   (omitting the unphysical points $\{\infty\}\times \bS^2$). This is because
   $\lambda$ enjoys the very inverse temperature $2\pi$ referred to $-\cD$.
On the other hand there are physical reasons to reject that candidate. 
 Indeed,  the circle $\bS^1= \bR \cup \{\infty\}$ admits two 
physically distinguishable
 points: The point at infinity, which cannot be reached physically because  it corresponds
 to a surface which does not belong to the Kruskal manifold.
  The other  
point corresponds to the bifurcation manifold where $\xi$ vanishes.  
(In the general case $\bM:= \bS^1\times \Sigma$  considered in this work,
$\bM$ itself cannot represent 
 a portion of spacetime due to the presence of closed causal curves lying in $\bS^1$ and thus one
 point of $\bS^1$ at least must be removed to make contact with physics.)
 The remaining points of $\bS^1$ are physically equivalent barring the fact that 
 they are either in the past or in the future of $\theta=0$. This determines 
two  regions  $\bF_- \equiv (-\pi,0)\times \bS^2$ and $\bF_+ \equiv (0,\pi)\times \bS^2$
 in the physical part $\bR \times \bS^1$,  of the manifold $\bS^1 \times \bS^2$, corresponding to respectively the   future and past 
 part -- with respect to the bifurcation manifold -- of the future Killing horizon of Kruskal spacetime. 
 Conversely, the  whole $PSL(2,\bR)$ unitary  representation, referred to the Fock space 
 $\gF_+(\cH)$ built upon $\lambda$, which, in turn, is invariant under the whole representation $U$,
 cannot select those physical regions. In particular $PSL(2,\bR)$ includes
 arbitrary displacements of the coordinate $\theta$. Those transformations connect 
 the physical regions with the points at infinity. For these reasons $\lambda$ seems not to be
completely  satisfactory from the point of view of physics in spite of its relevant thermal properties.\\
  Once a reference state $\mu$ is fixed on $\cW(\bM)$, the physical regions $\bF_\pm$
   correspond to von Neumann algebras $\cA(\bF_+)$ and   $\cA(\bF_-)$ 
   (based upon the GNS representation of $\mu$)   representing the observables
    in those regions.\\
   In the following we show that it is possible to single out
  those physical regions at {\em quantum,  i.e. Hilbert space,  level} through a sort of {\em spontaneous breaking of $SL(2,\bR)$ symmetry} referring to a new state $\lambda_\zeta \neq \lambda$ which {\em preserves} the relevant thermal properties.
We mean that the  following facts, actually valid for any manifold $\bM = \bS^1 \times \Sigma$, hold true. 
At algebraic level there is a representation $\alpha$ 
of M\"obius group of the circle $PSL(2,\bR)$ made of $^*$-automorphisms of the Weyl algebra 
$\cW(\bM)$.  
Moreover, we have seen in theorems \ref{1} and \ref{2} that there is a state $\lambda$
on $\cW(\bM)$  which is invariant under $\alpha$ and,  in the  GNS representation of $\lambda$, $\alpha$ is implemented 
unitarily and covariantly by a  representation $U$ of $PSL(2,\bR)$.
 We show below  that  there are other, unitarily inequivalent, GNS representations of $\cW(\bM)$
 based on new states $\lambda_\zeta$ which are
 no longer invariant under the whole $\alpha$, but such that, the residual
 symmetry is still covariantly and unitarily implementable and  singles out the algebras  $\cA(\bF_+)$ and   $\cA(\bF_-)$ as unique invariant algebras. We show also that every $\lambda_\zeta$  enjoys the same thermal (KMS) properties as $\lambda$ and it represents
a different thermodynamical phase with respect to $\lambda$.\\

\ssa{Symmetry breaking} We need  some definitions to go on. 
Coming back to the general case 
$\bM = \bS^1 \times \Sigma$ where $\Sigma$ 
is any Riemannian manifold, 
fix a standard frame $\theta\in (-\pi,+\pi)$ on $\bS^1$. 
The regions $\bF_\pm$ are defined as those containing the points 
$(0,\pi)\times \Sigma$
and $(-\pi,0)\times \Sigma$ respectively.
Consider the one-parameter subgroup of M\"obius transformations 
 $\bR \ni t\to \exp(t\cD)$ where ${\cal D}:= 
 -\sin\theta \frac{\partial}{\partial \theta}$ in $\bM$. 
 It admits $0$ and $\pi$ as unique fixed points. On the other hand, it is simply 
proved that (up to nonvanishing  factors)  ${\cal D}$ is the unique nonzero 
vector field in the representation of  $sl(2,\bR)$ which vanishes at  $0$ and $\pi$.
As a consequence that  subgroup
is the unique (up to rescaling of the parameter) nontrivial one-parameter 
subgroup of $PSL(2,\bR)$ which admits $(0,\pi)$ and $(-\pi, 0)$ as invariant 
segments.  The origin of the parameter $v$ of the integral curves of $-\cD$ can be arranged in order that
\beq
v = \Gamma(\theta):= \ln \left|\tan \frac{\theta}{2}\right| \label{v} \:,
\eeq 
where  $v$ ranges monotonically in $\bR$ with $dv/d\theta>0$ for  $\theta \in (0,\pi)$, 
whereas it ranges monotonically in  $\bR$ with $dv/d\theta<0$
for $\theta \in (-\pi,0)$.
In spite of its singularity at $\theta=0$, the function 
$\Gamma$ in (\ref{v})  is locally integrable. 
Thus  
for any fixed  function  $\zeta\in L^2(\Sigma,\omega_\Sigma)$, 
$\Lambda_\zeta(V(\omega)) := \lambda(V(\omega))  
e^{i\int_\bM \Gamma(\zeta \omega_+ + \overline{\zeta \omega_+})}$
is well defined if  $\omega\in \cD(\bM)$.
 Let us show that $\Lambda_\zeta$
extends to a state on $\cW(\bM)$.
It holds  $\Lambda_\zeta(V(0))=1$. Using (V1), (V2) and imposing linearity, $\Lambda_\zeta$ 
defines a linear  functional on the $^*$-algebra generated by all of objects $V(\omega)$.
 As $\lambda$ is positive, $\Lambda_\zeta$ turns out to be positive too, finally $\bR\ni t\mapsto \Lambda_\zeta(V(\omega))$ is continuous. 
For known theorems \cite{Lewis} there is a unique extension $\lambda_\zeta$ of  $\Lambda_\zeta$ to a state on
$\cW(\bM)$: 
If the real function $\zeta\in \loc$ is fixed, it is the unique state satisfying,   
\beq
\lambda_\zeta(V(\omega)) = \lambda(V(\omega)) 
\:e^{i \int_\bM  \Gamma  (\zeta\omega_+ + \overline{\zeta\omega_+})}
\:\label{GNS'}  
\eeq
for all $\omega \in \cD(\bM)$. 
Similar states, obtained by linear deformation of the vacuum state of a Fock representation
of Weyl algebra, are known in the literature as {\em coherent states}. 
They were studied in \cite{Roe70}  for photons in flat spacetime
and in \cite{Manuceau} (see also \cite{RST}). Several propositions presented in those work could be re-adapted to 
our case with some efforts.  We think anyway that the shortest way consists of giving independent proofs based on more modern general results of local quantum physics  \cite{Haag} as the proofs
of our propositions  are not very complicated. Similar states for free QFT defined in globally hyperbolic spacetimes 
containing a bifurcate Killing horizon give rise to the failure of the uniqueness property proved in \cite{KW} (see the first footnote 
on p. 70 in \cite{KW}).\\
$\lambda_\zeta$ and its GNS triple $(\gH_\zeta, \Pi_\zeta, \Psi_\zeta)$ enjoy the  remarkable properties
stated in the theorems below. \\

\teorema\label{4} {\em Fix a standard  frame $\theta$ on $\bS^1$ of $\bM = \bS^1\times \Sigma$,
  define  $\cD$ as in (\ref{vectors}) and the group of $^*$-automorphisms  $\alpha$ representing $PSL(2,\bR)$ as in  3.1,
 $\{\alpha^{(\cX)}_t\}_{t\in \bR}$ being  any one-parameter subgroup associated with
 the vector field  $\cX$.
If $\zeta\in L^2(\Sigma,\omega_\Sigma)$ and  $\lambda_\zeta$  is the state  defined in (\ref{GNS'})
with  GNS triple $(\gH_\zeta, \Pi_\zeta, \Psi_\zeta)$, the following holds: \\
{\bf (a)}   The map $V(\omega) \mapsto V(\omega) \:e^{i \int_\bM  \Gamma  (\zeta\omega_+ + \overline{\zeta\omega_+})}$, $\omega\in \cD(\bM)$, uniquely
extends to a $^*$-automorphism  $\gamma_\zeta$ on  $\cW(\bM)$
and
\beq
\lambda_\zeta(w) &=& \lambda(\gamma_\zeta w) \:,\:\:\:\:\: \mbox{for all $w\in \cW(\bM)$}\label{as1}\:,\\
\gamma_\zeta \circ \alpha_t^{(\cD)} &=& \alpha_t^{(\cD)} \circ \gamma_\zeta \:,\:\:\:\:\: \mbox{for all $t\in \bR$,} \label{as2}
\eeq
 {\bf (b)} (i) $\lambda_\zeta$ is  pure, (ii) if $\zeta \neq \zeta'$ a.e.,
 $\lambda_\zeta$ and $\lambda_{\zeta'}$ are not quasiequivalent,
 (iii) $\lambda_\zeta$ is invariant under 
$\{\alpha^{(\cD)}_t\}_{t\in \bR}$, but  it is  not under any
other one-parameter subgroup of $\alpha$ (barring those associated with $c\cD$ for $c\in \bR$ constant) when $\zeta \neq 0$ almost everywhere. \\
{\bf (c)}  $\gH_\zeta$ identifies with a Fock space $\gF_+(\cH_\zeta)$ with  vacuum vector $\Psi_\zeta$ and, for all $\omega \in \cD(\bM)$,
 \beq
\Pi_\zeta : V(\omega) \mapsto \hat{V}_\zeta(\omega) := e^{i\overline{\hat{\phi}_\zeta(\omega)}}\:, \:\:\:\:\: \mbox{where\:\:\: 
$\hat{\phi}_\zeta(\omega) := \hat{\phi}_0(\omega) + \left\{\int_\bM  \Gamma  (\zeta\omega_+ + \overline{\zeta\omega_+})\right\} I$,} \label{b}
\eeq
 $\hat{\phi}_0(\omega)$ being here the standard field operator in the Fock space $\gF_+(\cH_\zeta)$ as in 2.4.  \\
 {\bf (d)} There is a strongly continuous  one-parameter 
 group of unitary operators $\{U_\zeta^{(\cD)}(t)\}_{t\in \bR}$ with
\beq
\alpha_t^{(\cD)}(w) = U_\zeta^{(\cD)}(t)\: w\:U_\zeta^{(\cD)\dagger}(t) \label{reduced}
\:\:\:\:\:\:\mbox{for all $t\in \bR$ and $w\in\hat{\cal W}_\zeta(\bM):= \Pi_\zeta(\cW_\zeta(\bM))$}.
\eeq
Moreover (the derivative is performed in the strong sense where it exists)
\beq
\frac{d}{dt}|_{t=0}U_\zeta^{(\cD)}(t) = \frac{-i}{2} \overline{:\spa \Omega(\hat{\phi}_0, \cD\hat{\phi}_0)
\spa:} \label{idD}\:.\eeq}\\

\noindent The proof is in the appendix.\\

\teorema\label{5} {\em In the hypotheses of  theorem \ref{4}
  the following holds for  net of von Neumann algebras  \beq
{\cal A}_\zeta = \{{\cal A}_\zeta(I)\:\:|\:\: I\in {\cal I}\} \:,\:\:\:\: \mbox{with}\:\:\:\:
{\cal A}_\zeta(I)=\{\hat{V}_\zeta(\omega)\:\:|\:\: supp\: \omega \subset I\times \Sigma\}'' \label{Adelta}\:.
\eeq 
{\bf (a)}   ${\cal A}_\zeta\supset \hat{\cal W}_\zeta(\bM)$ and it  enjoys the following properties: (i) isotony, (ii) locality, (iii) $\{\exp(t\cD)\}_{t\in \bR}$-covariance, (iv) $U_\zeta^{(\cD)}$-invariance and uniqueness  of the vacuum
$\Psi_\zeta$, (v) cyclicity of the vacuum $\Psi_\zeta$, (vi) Reeh-Schlieder,
(vii) Haag duality, (viii) factoriality,
(iix) irreducibility, (ix) additivity.\\
{\bf (b)} If $\zeta\neq 0$ a.e.,
$\cA_\zeta(\bF_+):= \cA_\zeta((0,\pi))$ and $\cA_\zeta(\bF_-):= \cA_\zeta((-\pi,0))$ are the unique 
$\{U^{(\cD)}_t\}_{t\in \bR}$-invariant  algebras in ${\cal A}_\zeta$.\\
{\bf (c)}  If $\Delta$  is the modular operator associated with $\cA_\zeta(\bF_+)$ then
\beq
\Delta^{it} = U^{(\cD)}_{\zeta}(2\pi t)\:,\:\:\:\:\mbox{for all $t\in \bR$}\:. 
\eeq
Thus  $\lambda_\zeta$ is a KMS state on $\cA_\zeta(\bF_+)$
with temperature $T =1/2\pi$, with respect to 
$\{\alpha^{(-\cD)}_t\}_{t\in \bR}$ (extended to
$\sigma$-weak one-parameter group of $^*$-automorphisms of $\cA_\zeta(\bF_+)$ through (\ref{reduced})).}\\

\noindent{\em Proof}.
 (a) and (c) Since the difference between $\hat{V}_\zeta(\omega)$ and $e^{i\overline{\hat{\phi}_0(\omega)}}$
 amounts to a phase only,
 each algebra $\cA_\zeta(I)$ of  $\cA_\zeta$ coincides with the analog 
 constructed starting from operators $e^{i\overline{\hat{\phi}_0(\omega)}}$ and using the same $I\in {\cal I}$.
 Hence theorem \ref{3}
 and subsequent remark 2 hold using the field 
 $\hat{\phi_0}$, replacing $\Psi$ with $\Psi_\zeta$ and 
 employing the representation $U$ of $PSL(2,\bR)$ which leaves $\Psi_\zeta$ unchanged.
 Notice that $U$ does not implement  $\alpha$!
  In this way all the properties  cited in the thesis turn out to be automatically proved
 with the exception of (iii) and (iv). However using (\ref{as2}),  (\ref{idD})
 and (d) of theorem \ref{1} also those properties can be immediately proved. 
 The proof of (c) is straightforward. $\cA_\zeta((0,\pi))$ 
 coincides with the analog 
 constructed starting from operators $e^{i\overline{\hat{\phi}_0(\omega)}}$. In that case the thesis 
holds with respect to  the subgroup of $U$,  $e^{t\overline{:\Omega(\hat{\phi}_0,\cD(\hat{\phi}_0)):/2}}$
 (remark (2) after theorem \ref{3}). Now (\ref{idD}) implies the validity of the thesis in our case. \\
  (b)  Since $\cD$ admits the only zeros at $\theta=0$ and $\theta=\pi\equiv -\pi$, the only open 
nonempty and nondense intervals of $\bS^1$ which are invariant under the one-parameter group
$\{g^{(\cD)}_t\}_{t\in \bR}$ generated by $\cD$ are $(0,\pi)$ and $(-\pi,0)$.  
$\cD$-covariance reads
$U_\zeta^{(\cD)}(t) \cA_\zeta(I) U_\zeta^{(\cD)\dagger}(t) = \cA_\zeta(g^{(\cD)}_t(I))$ and thus
$\cA_\zeta((0,\pi))$ and $\cA_\zeta((-\pi,0))$ are invariant under $\{U_\zeta^{(\cD)}(t)\}_{t\in \bR}$.
Let us prove their uniqueness. Consider  the case 
of $I = (a,b)$ with $0\leq a<b<\pi$. There are $t'>0$ and $a'>0$, with $a'<b$ and
such that $g^{(\cD)}_{t'}(a',b) \cap (a,b) = \emptyset$. Therefore, by locality it holds
$[U_\zeta^{(\cD)}(t')\cA_\zeta((a',b))U_\zeta^{(\cD)\dagger}(t'), \cA_\zeta((a,b))]=0$,
i.e.
$[\cA_\zeta((a',b)), U_\zeta^{(\cD)}(-t')\cA_\zeta((a,b))U_\zeta^{(\cD)\dagger}(-t')]=0$. 
If  $\cA_\zeta((a,b))$ were invariant under $\{U_\zeta^{(\cD)}(t)\}_{t\in \bR}$, the latter identity above
would imply that 
$[\cA_\zeta((a',b)), \cA_\zeta((a,b))]=0$, and thus in particular
$\cA_\zeta((a',b)) \subset \cA_\zeta((a',b))'$ which is trivially false 
 because elements $\hat{V}_{\zeta}(\omega) \in  \cA_\zeta((a',b))$ generally do not commute.
 All the remaining cases can be reduced to that studied above with obvious adaptations. $\Box$\\

\noindent{\bf Remarks}. \\
{\bf (1)} (c) in the last theorem  is valid also replacing 
$\bF_-$ for $\bF_+$ and $\cD$ for $-\cD$ as well.
  Theorems \ref{4} and \ref{5}  hold in particular for  $\Sigma=\bS^2$ and $\bM=\bS^1\times \bS^2$. 
  In that case one finds easily that:\\  
{\em $\lambda_\zeta$ is  invariant under  the group of $^*$-automorphisms 
induced by the  action of $SO(3)$ as isometry group on  $\bS^2$
if and only if $\zeta$ is  constant  a.e. on $\bS^2$.}\\
Generic $\Sigma$ do not admit $SO(3)$ as group of isometries, in that 
case  $\lambda_\zeta$ is invariant under  the relevant isometry group 
of $\Sigma$ provided $\zeta$ is so.
Finally we notice that the hypotheses $\zeta\in L^2(\Sigma,\omega_\Sigma)$ can be relaxed in  
$\zeta\in L^1_{loc}(\Sigma,\omega_\Sigma)$ 
(the space of locally integrable functions on $\Sigma$ with respect to $\omega_\Sigma$)
both in the theorems \ref{4} and \ref{5}, the only 
result that could fail to hold is (ii) in (b) of the theorem \ref{4}.  \\
{\bf (2)} The theorems \ref{4} and  \ref{5} 
 refer to the pair of  segments $(0,\pi)$ and $(-\pi, 0)$ 
in the circle 
realized as the segment $[-\pi, \pi]$ with $-\pi\equiv \pi$. From a physical point of view 
there is no way to distinguish between the pair of regions $(0,\pi)$, $(-\pi, 0)$ and any other
pair of  open nonempty segments $I,J\subset \bS^1$  such that  $J= int(\bS^1 \setminus I)$. 
This is because there is no way to measure segments on $\bS^1$
as the metric is degenerate therein. In fact
the theorem can be stated for any  pair of such segments. To prove it we notice that  there 
exists a M\"obius diffeomorphism  $g: \bS^1\to \bS^1$ with  $I=g((0,\pi))$ and
$J= g((-\pi,0))$\footnote{Assume that, in coordinates $\theta$, $I$ has length 
equal or shorter than $J$.
The diffeomorphism $g^{-1}$ is the composition of a rigid rotation generated by $\cK$
which maps the center of $I$ in $0$, a dilatation generated by $\cD$ 
which enlarges the transformed  $I$ up to $(-\pi/2,\pi/2)$
and another anti-clockwise  rigid rotation of $\pi/2$.}. 
Hence, theorems \ref{4} and \ref{5} can be re-stated replacing $(0,\pi)$ and $(-\pi,0)$ with, respectively
$I$ and $J$, replacing the state  (\ref{GNS'}) 
with the state and  assuming to have fixed some $\zeta\in L^2(\Sigma,\omega_\Sigma)$,
$$\lambda_{I}(V(\omega)) := \lambda(V(\omega)) e^{i\Gamma_I(\omega)},\:\:\:\:\: 
\mbox{with $\Gamma_I(\omega):=\int_{\bM}\Gamma (\zeta  \:g^* \omega_+ + 
\overline{\zeta  \:g^* \omega_+})$}$$
and  replacing $\cD$ with the generator $\cD_I$ of the  one-parameter 
 subgroup   of $PSL(2,\bR)$
$\bR\ni t\mapsto\exp(t\cD_I) := g\circ \exp(t\cD) \circ g^{-1}$ which leaves invariant 
$I$ and $J$ ($\cD_I$ does not depend on the choice of $g$).\\
Notice also that  if $I,J$ is a pair of segments as said above and $h$ is any 
M\"obius transformation,
$h(I),h(J)$ still is a pair of open nonempty segments with $h(J) = 
int(\bS^1\setminus h(I))$ and it holds (using also lemma \ref{lemma1})
$$\lambda_{h(I)}(V(\omega)) = \lambda_I(V(h^*\omega))\:.$$
This fact means that the $PSL(2,\bR)$ symmetry, broken at Hilbert-space 
 level, is restored at algebraic level by considering the whole class of states 
 $\lambda_I$.
The residual Virasoro representation after breaking $PSL(2,\bR)$ symmetry is analyzed in the Appendix.\\
{\bf (3)} Considering again the particular case of the Kruskal manifold, the requirement (\ref{relation}),
that is
$-\sin \theta \partial_\theta = -\kappa^{-1} \xi\rest_\bF$, 
fixes the standard frame only 
up to a coordinate transformation $\theta' = \theta'(\theta)$, where $\theta'$ being any other
positive oriented  coordinate frame on $\bS^1$ satisfying $\sin \theta' \partial_{\theta'} =
\sin \theta\partial_\theta$.
Since our construction of quantum field theory on $\bM$ 
relies upon the choice of a standard  frame on $\bS^1$, a natural question is:
{\em Are quantum field theories based on $\lambda_\zeta$
and  its analog $\lambda'_\zeta$ with obvious notation,  unitarily equivalent?}
(Notice that $\zeta$ is the same for both states).
The answer is strongly positive because of the following general result.\\

\teorema\label{6} {\em With the same hypotheses as in theorem \ref{4},
let $\theta'$ be another standard frame on $\bS^1$. 
Referring to the coordinate frame $\theta'$, let $\cD'$ be the vector field analog of $\cD$
and let $\lambda'_\zeta$ be the 
 state analog of $\lambda_\zeta$ (both states defined on $\cW(\bM)$).
If $\cD'=\cD$ then, for any $\zeta\in L^2(\Sigma,\omega_\Sigma)$,
\beq \lambda'_\zeta = \lambda_\zeta\:.\eeq}

\noindent {\em Proof}. 
In our hypotheses 
$\theta'(\theta) = 2 \tan^{-1}(e^c \tan(\theta/2))$ for some $c\in \bR$.
The transformation  $\theta \to \theta'(\theta)$ {\em  
interpreted as an active diffeomorphism} is nothing but the action 
of the  element $\alpha_{-c}^{(\cD)}$ of the one-parameter group generated by $\cD$.
Since $\lambda_\zeta$ is invariant under that group ((b) in theorem \ref{4}) the 
thesis is  true.
$\Box$

\section{Towards  physical interpretations.}
Consider the case of $\bM$ constructed by the future Killing horizon of Kruskal manifold
(however theorem \ref{7} below  holds true for a generic degenerate manifold $\bM=\bS^1\times \Sigma$). 
As is well known the complete maximal Kruskal solution of Einstein equation 
describes a spacetime with an eternal pair of balck hole - white hole. 
However, some features (e.g. Hawking radiation) of real black holes produced by collapse 
can be modelled by using the right Schwarzschild wedge and the region containing the future singularity 
in Kruskal manifold,  the region about $\bF_+$ (see \cite{Wald0,Wald}) in particular. $\bF_+$ itself can be considered 
as (an extension of the) actual event horizon of a physical black hole.
The spacetime of a physical black hole obtained by stellar collapse has no white hole neither Killing bifurcate horizon. 
Nevertheless, in the sense stated below a physical black hole will asymptotically approach such a spacetime
(at least a spacetime including a bifurcate Killing horizon). 
Indeed, in \cite{RW} Racz and Wald considered a globally hyperbolic, stationary spacetime containing a black hole 
but no white hole, assuming,  further, that the event horizon $\bE$ of the black hole is a Killing horizon with 
compact cross-sections.
 With those hypotheses they proved  that, if surface gravity is non-zero and constant throughout 
the horizon, one can globally extend the initial spacetime so that the image of $\bE$ is a proper subset of a regular 
bifurcate Killing horizon in the enlarged spacetime. 
In that paper they also provided  necessary and sufficient conditions  for the extendibility of matter fields 
to the enlarged spacetime. These results support the view that any spacetime representing the asymptotic 
final state of a black hole formed by gravitational collapse may be assumed to possess a bifurcate Killing 
horizon (see \cite{RW} for details). 
Therefore, from a physical point of view,  it is worth investigating the physical meaning for the theory 
referred to the GNS representation of $\lambda_\zeta$ when restricting to the region $\bF_+$.\\

\ssa{Extremal KMS states: Existence of different thermodynamical  phases} 
By construction 
$\lambda_\zeta$ are KMS states on the $C^*$-algebra $\cW(\bF_+)$,
the  Weyl  algebra generated by  Weyl operators  $V(\omega)$ with $supp\: \omega \subset \bF_+$ which is contained in $\cA_\zeta(\bF_+)$.
As states on $\cW(\bF_+)$, $\lambda_\zeta$ and $\lambda_{\zeta'}$ can be compared
also if $\zeta\neq \zeta'$
(they do not belong to a common folium if (ii) in (b) of theorem \ref{4} holds, so they cannot
 be compared on a common von Neumann algebra of observables in that case).
The next theorem,  valid for the general case $\bM=\bS^1\times \Sigma$, shows that $\{\lambda_\zeta\}_{\zeta\in L^2(\Sigma,\omega_\Sigma)}$ is  a family of  extremal 
states in the convex space of $KMS$ states over $\cW(\bF_+)$ at inverse temperature $2\pi$
with respect to $-\cD$.\\

\teorema\label{7} {\em With the same hypotheses as in theorem \ref{4} the following holds.\\
{\bf (a)} Any state  $\lambda_{\zeta}$ 
(with $\zeta\in L^2(\Sigma,\omega_\Sigma)$) defines an  extremal states
in the convex set of $KMS$ states on the $C^*$-algebra $\cW(\bF_+)$ at inverse temperature $2\pi$ with respect 
to $\{\alpha^{(-\cD)}_t\}_{t\in \bR}$. \\
{\bf (b)} Different  choices of $\zeta$ individuate different states
on $\cW(\bF_+)$ which are not unitarily equivalent as well.}\\

\noindent{\em Proof}. 
Let   $(\gH_\zeta, \Pi_\zeta, \Psi_\zeta)$ be the GNS representations of  
$\lambda_\zeta$. The GNS  representations of  $\lambda_\zeta\spa\rest_{\cW(\bF_+)}$
must be (up to unitary equivalences) $(\gH_\zeta, \Pi_\zeta\spa\rest_{\cW(\bF_+)}, \Psi_\zeta)$ due to 
Reeh-Schlieder property ((a) in 
theorem \ref{5}) of 
 $\cA_\zeta(\bF_+)$. Since $\cA_\zeta(\bF_+)= \Pi_\zeta\spa\rest_{\cW(\bF_+)}''$ is a
  (type $I\spa I\spa I_1$) factor, the state $\lambda_\zeta\spa\rest_{\cW(\bF_+)}$
-- namely  $\Pi_\zeta\spa\rest_{\cW(\bF_+)}$ -- is primary
(see III.2.2 in \cite{Haag}). As a consequence, by  theorem 1.5.1 in \cite{Haag},
the KMS state  $\lambda_\zeta\spa\rest_{\cW(\bF_+)}$ is extremal in the space 
of KMS states on $\cW(\bF_+)$ with respect to $\alpha^{(-\cD)}_{t}$ at the temperature of $\lambda_\zeta\spa\rest_{\cW(\bF_+)}$
itself. Obviously  $\lambda_\zeta\spa\rest_{\cW(\bF_+)} \neq \lambda_{\zeta'}\spa\rest_{\cW(\bF_+)}$ 
 because, if $\zeta-\zeta'$ is not zero almost everywhere,  the integrals in the exponentials  defining
$\lambda_\zeta$ and $\lambda_{\zeta'}$  produce different 
results when applied to $V(\omega)$ with $supp\: \omega \subset \bF_+$ with a suitable choice of
$\omega$. The proof of non equivalence is the same as done (see the appendix) for the states defined in the whole von Neumann algebras. $\Box$\\

\noindent  The natural interpretation 
of this fact is that the states  $\lambda_\zeta$, restricted to the observables 
in the physical region $\bF_+$, are nothing but {\em different thermodynamical  phases}
of the same system at the Hawking temperature (see V.1.5 in \cite{Haag}).\\

\ssa{Bose-Einstein condensate and states $\lambda_\zeta$ with $\zeta$ real}
 In the following we assume that $\zeta$ is real.
Let us examine some features of the generators $\hat{\phi}_\zeta$ of the Weyl  representation associated with  $\lambda_{\zeta}$
when restricted to the physical region $\bF_+$.
Consider $\omega\in \cD(\bM)$ such that $supp\: \omega \subset  \bF_+$
and such that $\omega(v,s)$ can be rewritten as $\frac{\partial \psi(v,s)}{\partial v} dv\wedge \omega_\Sigma$
where $\psi$ is smooth and compactly supported in $\bF_+$. 
Similar ``wavefunctions'' $\psi$ have been considered in \cite{mopi3} building up 
scalar QFT on a Killing horizon ($\bF_+$ in our case).
 Using (\ref{v}) we can write the formal expansion
\beq
\hat{\phi}_\zeta(\omega) =  \int_{\bF_+}  \hat{\phi}_0(\theta_+(v)) \omega(v,s)  +    
\int_{\bF_+} \zeta(s)\:v \omega(v,s)\:. \label{dec}\eeq
In terms of wavefunctions, if $\Omega_{\bF_+}$ is the restriction of the right-hand side 
of the definition of $\Omega$ given in (\ref{cf})
to real  smooth functions compactly supported in $\bF_+$, it holds
\beq
\Omega(\psi, \hat{\phi}_\zeta) = \Omega_{\bF_+}(\psi, \hat{\phi}_0)  -  
 \int_{\Sigma}\left(  \int_{-\infty}^{+\infty} \psi(v,s) dv \right) \zeta(s)\omega_\Sigma(s) \:. \label{dec1}\eeq
The group of  elements $e^{itH_\zeta}:= U_\zeta^{(-\cD)}(t)$, $t\in \bR$ generates displacements $v\mapsto v-t$ in the variable $v$
in the argument of the wavefunctions $\psi$, since $v$ is just the parameter of the integral curves of $-\cD$ which takes the form
$\frac{\partial}{\partial v}$ in $\bF_+$. 
 Using Fourier transformation with respect to $v$ we can write down
\beq
\psi(v,s) = \frac{1}{\sqrt{2\pi}}\int_{\bR_+} dE \:\: \widetilde{\psi_+(E,s)} e^{-iEv} + \overline{\widetilde{\psi_+(E,s)}} 
e^{iEv}\label{dec2}\:.
\eeq
In  heuristic  sense $H_\zeta$ acts on the wavefunctions $\psi$
as the  multiplicative operator $\widetilde{\psi_+(E,s)} \mapsto E \widetilde{\psi_+(E,s)}$. 
Physically speaking, thermal properties of $\lambda_\zeta$ are referred just to the energy notion associated with that Hamiltonian.   Actually, as is well-known, this interpretation
must be handled with great care:  the interpretation of $\widetilde{\psi_+}$ as a representative of a one-particle quantum state can be 
done in a Fock space whose vacuum state does not coincide with the KMS state $\lambda_\zeta$
 (see V.1.4 and  the discussion in p. 219 of \cite{Haag}.)
Using (\ref{dec2}), (\ref{dec1}) can be re-written as
\beq
\Omega(\psi, \hat{\phi}_\zeta) = \Omega_{\bF_+}(\psi, \hat{\phi}_0)  -   
\sqrt{2\pi}\int_{\Sigma}
  \zeta(s)\widetilde{\psi(0,s)_+} \:\:\omega_\Sigma(s) \:. \label{dec3}
\eeq
 From (\ref{dec3}) it is apparent that $\hat{\phi}_\zeta$ gets  contributions from {\em zero-energy
modes} ($E=0$) as it happens in {\em Bose-Einstein condensate}. To this end  see chapter 6 of \cite{Popov}
 and 5.2.5 of \cite{BR2},  especially p. 72, where in 
the decomposition of the KMS state $\omega$ (after the thermodynamical limit) in $\bR^\nu$-ergodic states, the mathematical structure of the latter states 
resemble that of the states $\lambda_\zeta$. 
The decomposition of the field operator (\ref{dec}) into a ``quantum'' 
(with vanishing expectation value) and a ''classical'' (i.e. commuting with 
all the elements of the algebra) part is typical of the teoretical description of 
a boson system containing a Bose-Einstein condensate; 
the classical part plays the role of a {\em order  parameter} \cite{DGPS,Popov}.\\
Let us focus attention on the generator of $U_\zeta^{(-\cD)}(t)= e^{itH_\zeta}$ in the representation of a state 
$\lambda_\zeta$. 
Using theorem \ref{4} we find (both sides are supposed to be restricted to the core  $F(\cH_\zeta)$)
$$H_\zeta =\int_{\bM}  \sin(\theta) :\spa\frac{\partial \hat{\phi}_\zeta}{\partial \theta} 
\frac{\partial \hat{\phi}_\zeta}{\partial \theta} \spa:\spa (\theta,s)\: d\theta \wedge\omega_\Sigma(s)\:.$$
Indeed, if  $\theta_\pm(v) = \pm 2\tan^{-1}(e^v)$ are the inverse functions of 
$v= \Gamma(\theta)$ in $\bF_+$ and $\bF_-$ respectively, passing from coordinates $(\theta,s)$ 
to coordinates $(v,s)$ and employing the field $\hat{\phi}_0$ the right-hand side of the 
formula above can be rearranged as
 $$H_\zeta = \lim_{N\to +\infty}  \left\{\int_{\bF_+}   \chi_N(v) 
:\spa\frac{\partial\hat{\phi}_0}{\partial v}
\frac{\partial\hat{\phi}_0}{\partial v} \spa: \spa (\theta_+(v),s) \:dv\wedge\omega_\Sigma(s) +
||\zeta||^2 \int_{\bR}   \chi_N(v) dv\right.$$
$$- \left.\int_{\bF_-}   \chi_N(v) :\spa
\frac{\partial\hat{\phi}_0}{\partial v}\frac{\partial\hat{\phi}_0}{\partial v}\spa: \spa(\theta_-(v),s)\:
dv\wedge\omega_\Sigma(s) -
||\zeta||^2 \int_{\bR}   \chi_N(v) dv\right\}\:,$$
where the function $\chi_N$ is smooth with compact support in $[-N,N]$ and  becomes the constant 
function $1$ for $N\to +\infty$.
The two constant terms in brackets cancel out each other, they having the opposit sign, and the final form 
of $H_\zeta$ is just that in (d) of theorem \ref{4}.  The normal ordering prescription used in the 
integrals is defined by 
subtracting $\left(\Psi,\hat{\phi}_0 (\theta',s')\hat{\phi}_0 (\theta,s) \Psi \right)$ 
before applying derivatives
and then smoothing with a product of delta in $\theta,\theta'$ and $s,s'$. We do not enter into mathematical details here which are quite standard procedures of applied microlocal analysis
similar to that used in Hadamard regularization \cite{BrunettiFredenhagen, HollandsWald, morettistress}. \\
 From the decomposition of $H_\zeta$ written above, we see that it is made of two contributions
$H_\zeta^{(-)}$, $H_\zeta^{(+)}$ respectively localized 
at the two disjoint regions of $\bF$, $\bF_-$ and $\bF_+$. The two terms have  the same
value with opposit sign as one expects form  the indefiniteness of the  self-adjoint 
generator $D$ 
(corresponding to the fact that the Killing vector $-\cD$ 
changes orientation passing from $\bF_+$ to $\bF_-$).
Let us concentrate on the second term in the contribution $H_\zeta^{(+)}$ to $H_\zeta$ 
due to $\bF_+$. It is a volume divergence
 $$\lambda_\zeta(H^{(+)}_\zeta) = E_\zeta := ||\zeta||^2 \int_{\bR} dv\:.$$
 This can be interpreted as the energy of the BE condensate localized at $\bF_+$ whose density is {\em finite}
 and amounts to $||\zeta||^2$. \\
 
\ssa{Conclusions: Can the condensate describe physical propeties of a black hole?}
Here, to conclude,  we try to give some hints to relate the properties of the condensate with spacetime, i.e. Schwarzschild black hole, 
properties. To do  it we start from a deeper point of view. The only difference between 
two different Schwarzschild black holes concerns  their masses, that is their Schwarzschild radii.
Since we want to ascribe this difference to a feature of a state, the background and the system 
supporting the state
must be independent from the black-hole radius. In this way the states $\lambda_\zeta$ have to be referred to 
a quantum field theory on an abstract manifold $\bM = \bS^1\times \bS^2$ with a metric
on $\bS^2$ which does not coincide with the actual metric of a particular black hole. 
We assume the hypothesis of spherical symmetry so that the metric on $\bS^2$ is determined 
by fixing the value of an adimensional parameter only (the radius rate for instance). 
In this view a state $\lambda_\zeta$
on the  scalar field $\hat{\phi}$ must fix the geometry of the black hole
under the constraints of the presence of
a Killing horizon and spherical symmetry. 
Since we are in fact dealing with quantum gravity we adopt natural Planck units ($\hbar=c=G=1$)
so that we can emply pure numbers in the following. In particular, the pure number defining 
the radius of $\bS^2$ will be
denoted by $r_0$.\\ 
The idea that the assignment of a (classical) scalar field fixes the metric of a 
spacetime (solution of Einstein equations) when other constraints are given on the metric  
is not new, the so-called
dimensional-reduction theory  for gravitation leads to such a scenario (e.g. see \cite{piva04} with cited references)
where the scalar field is related to the dilaton field. Now we adopt a similar point of view 
but, in addition, we assume also that the assignment of the configuration of the scalar field
is due to the assignment of a quantum state of that field. 
Let us see how this idea can be implemented from 
the following remark.\\
Spherical symmetry implies that $\zeta$ must be constant on $\bS^2$ (see remark 1 after theorem 4.2).
Since the considered states are coherent the field admits a nonvanishing averaged value. Formally it holds 
\beq\lambda_\zeta( \hat{\phi}_\zeta(\theta,s)) = v_\zeta(\theta)\:.\label{prefcoord}\eeq
(See remarks below).
Hence  the mean value of $\hat{\phi}_\zeta$ with respect to $\lambda_\zeta$ picks out a preferred
coordinate frame along the light lines of $\bF_+$. 
 So, up to the choice of the origin, the mean value of the field $\hat{\phi}_\zeta$ 
 defines a {\em preferred coordinate} $v_\zeta$ in the physical region $\bF_+$. 
 Now the natural hypotheses is that $v_\zeta$ is the parameter of the Killing field $\xi\rest_{\bF_+}$
 of the considered black hole as in 4.1.
 In other words we are saying that $\zeta$ determines a black hole in the class of Schwarzschild ones 
 by determining its surface gravity through the identity (both sides are pure numbers since we are 
 employing natural Planck units):
 \beq\zeta = \kappa^{-1}\label{zetakappa}\:.\eeq
 Such a black hole must have horizon surface $S_\zeta= \pi \zeta^2$. 
 As a consequence we find that \beq||\zeta||^2 = 4\pi \zeta^2 r_0^2\eeq scales as the actual surface
  of the black hole horizon (and it is exactly the measure of the surface provided $r_0= 1/2$). This provides some clues for 
  an interpretation of 
  $||\zeta||^2$ that is, equivalently,  the {\em density of energy} of the condensate $E_\zeta/\int_\bR dv$. \\

 \noindent {\bf Remarks} A pair of mathematical remarks are necessary interpret (\ref{prefcoord}).\\ 
{\bf (1)} $\lambda_\zeta( \hat{\phi}_\zeta(\theta,s))$
is not well defined and it could be thought as the weak limit of a sequence 
 $\lambda_\zeta( \hat{\phi}_\zeta(\omega_n))$ where the forms $\omega_n$ regularize 
 Dirac' s delta centered in $(\theta,s)\in \bF$.\\
{\bf (2)} Furthermore, one  has to take into account that the allowable forms
have the shape  $\omega_n(\theta,s) = \frac{\partial f_n(\theta,s)}{\partial \theta} d\theta \wedge \omega_\Sigma$
were $f_n$  is {\em periodic} in $\theta$. It is not possible to produce a regularization sequence for
$\delta(s,s')\frac{\partial \delta(\theta'-\theta)}{\partial \theta} d\theta \wedge \omega_\Sigma$
in this way due to the periodic constraint.
The drawback can easily be skipped by fixing an origin $v_{\zeta0}$ for $v_\zeta$ 
(corresponding to some $\theta_0$)  for the coordinate $x$. In other words one considers 
a sequence of forms $\omega^{(\theta,s)}_n$ induced by smooth $\theta$-periodic functions
 $f^{(\theta,s)}_n(\theta')=\delta_n(s'-s)\left[\Theta_n(\theta-\theta')+\Theta_n{(\theta'-\theta_0)}\right]$, where
 $\{\delta_n(s')\}$ regularize $\delta(s')$ and $\{\Theta_n(\theta')\}$ regularize 
 the step distribution whose derivative is just $\delta(\theta')$. In this sense
 $$\lim_{n\to +\infty}  \lambda_\zeta( \hat{\phi}_\zeta(\omega^{(\theta,s)}_n)) = 
 v_\zeta(\theta) -v_{\zeta0}\:.$$

 The presented results could lead to an interesting scenario
which deserves future investigation. The Kruskal spacetime could 
be a classical object arising by spontaneous breaking of $SL(2,\bR)$ symmetry
as well as Bose-Einstein condensation due to  a state  of a local QFT defined on  a certain conformal net.
In particular the abstract field operator $\phi$ can be seen 
 as a noncommutative coordinate on $\bF_+$. (Obviously noncommutativity 
 arises from canonical commutation relations 
  $[\phi(\theta,s),\phi(\theta',s')] = iE(\theta,s,\theta',s')$.) Commutativity is restored under the choice of an 
 appropriate coherent state on that $^*$-algebra considering the averaged values of the field. This state fixes
 also the actual black hole. 
 (A recent  remarkable application of some ideas of 
 noncommutative geometry to conformal net theory and black holes appears in 
 \cite{Kawahigashi-Longo}.) With a pair of fields $\phi$
  defined on $\bF$ and the other defined on the past Killing horizon $\bP$ we may define,
 through the outlined way, global  null coordinates in the complete $r,t$ section of  right Schwarzschild wedge.  
A subject deserving future investigation concerns the issue if, in addition to the null coordinates in the plane $r,t$,  
 it is possible to give a quantum 
interpretation to the transverse coordinate and the whole metric of the Kruskal manifold.

\appendix
\section{Appendix}

\ssb{Fock representation and GNS theorem}
The interplay of the  Fock representation presented in Section 3  and GNS theorem \cite{Haag,BR} is simply sketched. 
Using notation introduced therein, if $\Pi  : {\cal W}(\bM) \to \hat{\cal W}(\bM) $ denotes the unique 
 ($\Omega$ being nondegenerate)
 $C^*$-algebra isomorphism between those two Weyl representations,
it turns out that $({\gF_+}(\cH ), \Pi ,  \Psi )$ 
is the GNS triple associated with a particular pure algebraic state $\lambda$
({\em quasifree} \cite{BR,KW}  and invariant under
the automorphism group associated with  $\partial_\theta$)
on ${\cal W}(\bM)$ we go to introduce.
Define $$\lambda (W(\psi)) := e^{-\langle \psi_+, \psi_+\rangle /2}$$
then extend $\lambda$ to the $^*$-algebra finitely generated by all the elements
$W(\psi)$ with $\psi \in \cS(\bM)$, by linearity  and using (W1), (W2). It is simply proved that, $\lambda(\bI)=1$ and
$\lambda(a^*a)\geq 0$ for every element $a$ of that $^*$-algebra so that $\lambda$
is a state. As the map $\bR\ni  t\mapsto\lambda(W(t\psi))$ is continuous,  
known theorems \cite{Lewis} imply that $\lambda$ extends uniquely to a state $\lambda$
on the complete Weyl algebra ${\cal W}(\bM)$.
On the other hand, by direct computation, one finds that $\lambda(W(\psi)) = \left\langle \Psi ,  \hat{W}(\psi)  \Psi  \right\rangle$.
 Since a state on a $C^*$ algebra is continuous, this relation can be extended to the whole algebras 
by linearity and continuity
and using  (W1), (W2) so that a  general GNS relation is verified:  
\beq\lambda(a) =
 \left\langle \Psi ,  \Pi (a) \Psi  \right\rangle
\:\:\:\:\:\mbox{for all $a\in {\cal W}(\bM)$}\:. \label{aa} 
\eeq 
To conclude, it is sufficient to show that  $ \Psi $
is cyclic with respect to $\Pi$. Let us show it.
If $\hat{\cal F} (\bM)$ denotes the $^*$-algebra
generated by field operators $\Omega(\psi,\hat{\phi})$, $\psi\in \cS(\bM)$, defined on $F(\cH)$,
$\hat{\cal F} (\bM) \Psi $
is dense in the Fock space (see proposition 5.2.3 in \cite{BR2}).  Let   $\Phi \in \gF_+(\cH)$  be a vector orthogonal to both $\Psi$ and to all the vectors
$\hat{W}(t_1\psi_1)\cdots \hat{W}(t_n\psi_n)\Psi$ for  $n=1,2,\ldots$ and $t_i\in \bR$ and $\psi_i\in \cS(\bM)$.
Using Stone theorem to differentiate in $t_i$ for $t_i=0$, starting from $i=n$ and proceeding backwardly up to $i=1$, one finds that  $\Phi$ must also be  orthogonal to all of the vectors
$\Omega(\psi_1,\hat{\phi})\cdots \Omega(\psi_n,\hat{\phi})\Psi$
and thus vanishes because $\hat{\cal F} (\bM) \Psi$
is dense.
This result means that  $\Pi ({\cal W}(\bM)) \Psi $
is dense in the Fock space too, i.e. $\Psi$ is cyclic
with respect to $\Pi $.  Since $\Psi$ satisfies also (\ref{aa}),  the uniqueness of the GNS triple 
 proves that the triple 
$({\gF_+}(\cH ), \Pi ,  \Psi )$ 
is just (up to unitary transformations) the GNS triple  associated with 
$\lambda$. Since the Fock representation is irreducible, $\lambda$ is pure.  \\

\ssb{Residual Virasoro representation after breaking $PSL(2,\bR)$ symmetry}
The complex Lie  algebra $(\frak{a}, \{\cdot,\cdot\}, \imath)$  of vector field on
$\bS^1$ (see discussion in 3.1) is made of vector fields on $\bS^1$ whose diffeomorphism groups,
generated by their real and imaginary parts, do not admit (in general)
$\bF_\pm$ as invariant regions, when extended to $\bM = \bS^1\times \Sigma$. This happens in particular for generators $\cL_n = i
e^{in\theta}\partial_\theta$. However, it is possible to rearrange that basis in order to partially
overcome
the problem. Consider the equivalent basis of $\frak{a}$ made of the following real  vector fields $-i\cL_0$, 
$\cE_n := (1- \cos((2n)\theta))\partial_\theta$, $\cO_n := (1+\cos((2n+1)\theta))\partial_\theta$,
$\cG_n := - \sin(n\theta)\partial_\theta$ with $n=1,2,\ldots$.
Barring $-i\cL_0$ and $\cO_n$, the other fields admit $\bF_\pm$ as invariant regions.
Moreover the fields $\cG_n$ define a Lie algebra with respect to the usual Lie bracket
whereas $\cE_n$, or $\cE_n$ together $\cG_n$, do not so. However allowing infinite linear
combinations of vector fields -- using for instance $L^2$-convergence
for the components of vector fields with respect to $\partial_\theta$ (the same result hold
anyway using stronger notions of convergence as uniform
convergence of functions and their derivatives up to some order) --
one sees that each  $\cE_n$ can be expanded as an infinite linear combination of $\cG_n$.
 From these considerations one might expect, at least, that fields $\cE_n$, but {\em not the vectors
$\cL_n$ and $\cO_n$}, admit some operator  representation in $\gH_\zeta$
in terms of the field operator $\hat{\phi}_\zeta$.
In fact this is the case if $\zeta$ is a real function in $L^2(\Si,d\Si)$.
If one tries to define  operators $L^{(c)}_{\zeta n}$ as
in (\ref{useful}) with $\hat{\phi}^{(c)}$ replaced
with $\hat{\phi}_\zeta^{(c)} := \hat{\phi}^{(c)} + \zeta  \Gamma$, one immediately faces
ill-definiteness of those operators due to infinite additive terms and
the same problem arises for formal operators $O_{n}^{(c)} :=L^{(c)}_{\zeta 0} +({L^{(c)}_{\zeta
2n+1}+L^{(c)}_{\zeta\: -2n-1}})/{2}$ and also for $E_{n}^{(c)} :=L^{(c)}_{\zeta 0}-({L^{(c)}_{\zeta
2n}+L^{(c)}_{\zeta\: -2n}})/{2}$.  However these terms cancel out if considering  the  operators
$G_{ n}^{(c)}:= ({L^{(c)}_{\zeta\: -n}-L^{(c)}_{\zeta n}})/{(2i)}$
with $n=1,2,\ldots$,
which are  well defined and essentially selfadjoint  on $F(\cH_\zeta)$.
Moreover, the operators $G_{ n}^{(c)}$ define a Lie  algebra
with respect to the commutator.
(Direct inspection shows that if $c=\infty$ none of the considered  operators  is well-defined on $F(H_\zeta)$.)
It is plausible that operators  $\overline{G_{n}^{(c)}}$
define one-parameter groups which implement covariance with respect to analogous groups
of diffeomorphisms generated by associated vector fields $\cG_n$, and that  the exponentiation of the algebra
of $G_{ n}^{(c)}$ produces a unitary representation of
a (perhaps the) subgroup of $Dif\spa f^+(\bS^1)$ of the diffeomorphisms which leaves $\bF_\pm$ invariant.
However, it is worth stressing that, barring the case $\overline{G_{1}^{(c)}}$ which generates just
$U_\zeta^{(\cD)}(t)$,
$\Psi_\zeta$  is not invariant under the remaining unitary groups.\\

\ssb{Proofs of some theorems}

\noindent {\em Proof of Proposition \ref{prop2.2}}.
Let  $\theta$ be a standard frame on $\bS^1$.
 Assume the condition (a) holds. We can write
 $\omega=\epsilon_f$ and $\omega'=\epsilon_{f'}$ for some functions 
 $f,f'\in \bC^\infty(\bS^1\times \Sigma;\bC)$. 
 To use these facts we notice that, in the general case, it holds
 $E(\epsilon_f,\epsilon_{f'})= \Omega(f,f')/4$ by proposition \ref{prop1}. Therefore, by (V2),
 to conclude the proof it is sufficient 
 to show that $\Omega(f,f')=0$. Let us prove it.
 In our hypotheses $f'$ is constant in the variable $\theta$ in $I\times \Sigma$ 
 since $\frac{\partial f'(\theta,s)}{\partial \theta}=0$ therein and $I\times \Sigma$ is  connected 
 by paths with $s$ constant. Moreover,  if $t,t'$ are the 
 endpoints of $I$, it must 
 hold $f(t,s)= f(t',s)$ for every $s\in \Sigma$. Indeed $\frac{\partial f(\theta,s)}{\partial \theta}=0$ 
 vanishes outside $I\times \Sigma$ -- and thus $f$ is constant in $\theta$ 
 in that set as before -- and $f$ is periodic in $\theta$ at $s$ fixed by hypotheses. 
  Integrating by parts in the right-hand side of the definition of $\Omega$ given in (\ref{cf}) 
  with $f$ and $f'$ in place of $\psi$ and $\psi'$,
$$\Omega(f,f')  =2\int_\Sigma \omega_\Sigma(s) 
\int_{\bS^1} f'(\theta,s) \frac{\partial f}{\partial\theta}(\theta,s)  d\theta\
= 2\int_\Sigma \omega_\Sigma(s) 
\int_I f'(\theta,s) \frac{\partial f}{\partial\theta}(\theta,s)  d\theta\:.$$
$f'$ is constant in $\theta$ in $I\times \Sigma$ and $f(t',s)= f(t,s)$, $t,t'$ being the extreme points of $I$, so that 
$$\frac{1}{2}\Omega(f,f') = \int_I f'(\theta,s) \frac{\partial f}{\partial\theta}(\theta,s)  d\theta=
f'(s)\int_I  \frac{\partial f}{\partial\theta}(\theta,s)  d\theta=
f'(s) (f(t',s)-f(t,s))=0\:.$$
Now suppose that (b) holds true. In this case one has
$$ i\Omega(f,f') =2\int_\Sigma \omega_\Sigma(s) 
\int_{\bS^1} f'(\theta,s) \frac{\partial f}{\partial\theta}(\theta,s)  d\theta\
= 2\int_S \omega_\Sigma(s) 
\int_{\bS^1} f'(\theta,s) \frac{\partial f}{\partial\theta}(\theta,s)  d\theta\:.$$
Since $\frac{\partial f'(\theta,s)}{\partial \theta}=0$ in the set $\bS^1 \times S$
which is connected by paths with $s$ constant, $f'$ does not depend on $\theta$
in that set and thus
$$\frac{1}{2}\Omega(f,f') =  2\int_S \omega_\Sigma(s)  f'(s) 
\int_{\bS^1} \frac{\partial f}{\partial\theta}(\theta,s)  d\theta =2\int_S \omega_\Sigma(s)  f'(s)  =0\:.$$
Finally (W2) or equivalently (V2) entails the thesis.  $\Box$\\

\noindent {\em Proof of Theorem \ref{1}}.
 The operator  $L:= K^2+S^2+D^2$ is essentially selfadjoint on $F(\cH )$
since the dense invariant space $F(\cH )$
 is made of  analytic vectors. The proof is straightforward by
 direct estimation of $||L^n \Psi||$ with $\Psi\in F(\cH )$ (there is a constant
 $C_\Psi \geq 0$ with $||L^n \Psi|| \leq C_\Psi^n$).
As a consequence of some results by Nelson
 (Theo. 5.2, Cor. 9.1, Lem. 9.1 and Lem. 5.1 in  \cite{Ne})  
 the Hermitean operators $iR (x)$ with $x\in sl(2,\bR)$  are essentially selfadjoint on $F(\cH )$
 and  there is a unique strongly-continuous representation  $\widetilde{SL(2,\bR)} \ni g\mapsto U (g) : \gF_+(\cH ) \to  \gF_+(\cH )$
such that (\ref{Ne}) holds true. \\
(a) $k$ generates the one-parameter subgroup $\bS^1$ in $SL(2,\bR)$ -- that is $\bR\ni t\mapsto \exp(tk)$ with period $4\pi$ --
as well as  the one-parameter subgroup $\bR \ni t\mapsto l(t)$  isomorphic to $\bR$ in $\widetilde{SL(2,\bR)}$.
 From the general theory of $ \widetilde{SL(2,\bR)}$
representations, a representation  $\widetilde{SL(2,\bR)} \ni g\mapsto V(g)$ is in fact
a representation of $SL(2,\bR)$ if  $t\mapsto V(l(t))$ has period $4\pi/k$ for some integer $k\neq 0$.
It is simply proved that the operator $K$ is
the tensorialization of the operator defined on $\ell^2(\bC)\otimes L^2(\Sigma,\omega_\Sigma)$
by extending 
$$\{C_n\}_{n=1,2,\cdots}\otimes u_j \mapsto \{nC_n\}_{n=1,2,\cdots}\otimes u_j $$
by linearity. As a consequence the spectrum of $\overline{K}$ is the set 
$\sigma(\overline{K})= \{0, 1,2,\ldots\}$ where the eigenspace with eigenvalue $0$
is one-dimensional and it is  generated by the vacuum state $ \Psi $. This implies that
$\bR \ni t\mapsto e^{it \overline{K}} = U (l(t))$ has period $2\pi$. As a first consequence
$U $ is a proper representation of $SL(2,\bR)$. Furthermore, since
 $\sigma(\overline{K})$ is nonnegative, the representation is a positive-energy 
 representation. Finally, notice that $-I = e^{2\pi k}$ and thus
 $U (-I)=  e^{i2\pi \overline{K}} = I$ and so $U $
 is a representation of $PSL(2,\bR) := SL(2,\bR)/\pm I$.\\
(b) and (c). 
 From direct inspection one  sees that the operators $K,S,D$ are tensorializations
 of the respective operators $K\spa\rest_{{\cal H} },
 S\spa\rest_{{\cal H} }, D\spa\rest_{{\cal H} }$, in particular
their restriction to the space generated by the vacuum vector coincide with the  operator $0$. 
 Moreover, decomposing  ${\cal H}  = \ell^2(\bC)\otimes L^2(\Sigma,\omega_\Sigma)$,
one finds
  $$K\spa\rest_{{\cal H} }=K_{0} \otimes 0\:,\:\:\:\:
 S\spa\rest_{{\cal H} } =S_{0} \otimes 0\:,\:\:\:\:
 D\spa\rest_{{\cal H} } =D_{0} \otimes 0\:, 
  $$
where $K_0, S_0, D_0$  are obtained  by restricting to the one-particle space
the operators $K,S,D$  defined in the case  $\bM= \bS^1$ (without transverse manifold).
 Using again Nelson results these operators give rise to a representation $\widetilde{SL(2,\bR)}\ni 
 g \mapsto V(g) \otimes I$ in ${\cal H} $.
 (This representation is, in fact, an irreducible representation of $SL(2,\bR)$, see \cite{mopi4}.)
 By tensorialization this representation extends to a representation $U'$
 in the whole Fock space. By construction, 
 the generators $iK',iS',iD'$ of this representation ad  associated with   $k,s,d$ respectively coincides with 
 $iK,iS,iD$ on $F({\cal H} )$ respectively. 
 Nelson's uniqueness property implies that
 $U'= U$. By construction $U$
 ($=U'$)
 admits every space with finite number  of particles as invariant space, including the space
with zero particles  spanned by
the vacuum state.\\
 (d)  First of all, as said above,  $U$ leaves invariant the space generated by the vacuum vector 
 $\Psi$ so that
it is an invariant vector up to a phase. Let us show that this is the only unit vector with this property.
By (b), the operator $\overline{D}$ is the tensorialization of $\overline{D_0}\oplus I = \overline{D_0\oplus I}$ where the generator of $V$,
$\overline{D_0}$, is defined on the one-particle space in the case of the absence of $\Sigma$, $\ell^2(\bC)$, and $I$ acts on $L^2(\Sigma,\omega_\Sigma)$. In \cite{mopi3,mopi4}  the representation $V$
has been studied, realized, under a suitable Hilbert space isomorphism,  in the space $L^2(\bR^+,dE)$. In that space $\overline{D_0}$ is the closure of the essentially-selfadjoint operator  $-i(Ed/dE +1/2)$. The original dense, invariant  domain of $-i(Ed/dE +1/2)$ is a core for $D_0$ 
made of smooth functions on $(0,+\infty)$ (see \cite{mopi3} for details) of the form
$\sqrt{E}e^{-\beta E}P(E)$ with $\beta>0$ a constant not depending on the considered function
and $P$ any polynomial.
Under the unitary transformation $U$, which takes the form  $(U\psi)(x) := (2\pi)^{-1/2}\int_{0}^{+\infty} e^{-ix\ln E}  \psi(E)/\sqrt{E} dE$ on the domain of $-i(Ed/dE +1/2)$, this operator 
becomes the operator position $X$ (i.e  $(X\psi)(x) = x\psi(x)$) on $L^2(\bR,dx)$
restricted to a core contained in the Schwartz space. As a consequence $\sigma(\overline{D_0})= \sigma_c(\overline{D_0}) = \sigma(X) =\bR$
and, similarly,  $\sigma(\overline{D_0 \oplus I})= \sigma_c(\overline{D_0 \oplus I}) =\bR$. Therefore, passing to the tensorialization,  $\sigma(\overline{D})= \bR$ and $\sigma_p(\overline{D}) = \{0\}$ with, up to phases, unique eigenvector given by the vacuum vector  $\Psi$.
If $\Phi$ is a unit vector which is up-to-phases invariant under $U$, it must be in particular
$e^{it\overline{X}} \Phi = u_X(t) \Phi$
 where $X$ is any real linear combination  of $K,S,D$
 and $|u_X|=1$. As the domain of $X$ is dense, it contains  a vector
 $\Phi'$ with $\langle \Phi', \Phi\rangle \neq 0$ and thus
 $u_X(t) = \langle e^{-it\overline{X}} \Phi', \Phi\rangle/\langle \Phi', \Phi\rangle$
 is differentiable at $t=0$ by Stones' theorem. As a consequence, the left-hand side
 $e^{it\overline{X}} \Phi = u_X(t) \Phi$ must be differentiable  at $t=0$. By Stone theorem $\Phi$ belongs to the domain of 
 $\overline{X}$ and it holds
$ \overline{X} \Phi = \lambda_X \Phi$
where $\lambda_X = -i du_X/dt|_{t=0}$. 
Specializing the identity to $X=D$, from the spectral structure of $\overline{D}$, one concludes that it must be $\lambda_D=0$ and,
up to phases,  
$\Phi= \Psi$.  $\Box$\\

\noindent{\em Proof of Theorem \ref{2}}. 
 (a) and (b). 
 To  establish  (\ref{a})  it is sufficient to prove those identities
 for $w=\hat{V}(\omega)$ 
 with $\omega \in \cD(\bM)$ and $g\in 
 PSL(2,\bR)$.
Actually, with the said choices for $w$ 
 \beq
U (g) \:a\: U^\dagger (g)
&=& \alpha'_g(a)\:,\:\:\:\: \mbox{for all $a\in \hat{\cal F} (\bM)$} \label{bx}\:.\eeq 
  implies (\ref{a}). 
  For if (\ref{bx})  holds, taking  the adjoint twice for both sides 
 one gets the relations for selfadjoint field operators
$U (g) \:\overline{\hat{\phi}(\omega)}\: U^\dagger (g)
= \overline{\hat{\phi}(\omega^{(g^{-1})})}$.
Then (\ref{utile}) implies  (\ref{a}) for $w=\hat{V}(\omega)$
via standard spectral theory. To conclude the proof of (a) it is now  sufficient to show the validity of (\ref{bx}) with $a=\hat{\phi}(\omega)$ or of the equivalent statement 
\beq
U (g) \:\Omega(\psi,\hat{\phi})\: U^\dagger (g)
= \Omega(\psi^{(g^{-1})}, \hat{\phi})\:,\:\:\:\: \mbox{for all $\psi \in \cS(\bM)$ and $g\in PSL(2,\bR)$} \label{b2}\:.
\eeq
In turn, using the fact that $U $ preserves the vacuum vector and is  the tensorialization of $U \spa\rest_{\cal H }$  (theorem \ref{1}) as well as (\ref{rfield}) one sees that  (\ref{b2}) is equivalent to
\beq
\psi^{(g)} = U (g^{-1})\spa\rest_{\cal H } \psi_+ + 
\overline{U \spa(g^{-1})\rest_{\cal H } \psi_+} 
 \:,\:\:\:\: \mbox{for all $\psi \in \cS(\bM)$ and $g\in PSL(2,\bR)$} \label{b3}\:.
\eeq
Let us prove (\ref{b3}). If $\psi\in \cS(\bM)$ and $g\in Dif\spa f^+(\bS^1)$ the map $\psi \mapsto \psi^{(g)}$ induces  a $\bR$-linear  map from the space of $\theta$-positive frequency
parts $\psi_+$ to the same space given by $$\psi_+ \mapsto 
S(g)\psi_+ :=
((\psi_+ +\overline{\psi_+})^{(g^{-1})})_+ .$$
In this way the action of $g$ on the wavefunction $\psi$ is equivalent to the action of $S(g)$
on its positive frequency part $\psi_+$:
\beq
\psi^{(g^{-1})} = S(g)\psi_+ + \overline{S(g)\psi_+} \label{center}\:.
\eeq
However,
in general,  $S(g)$ is not $\bC$-linear (and thus it  cannot be seen as a map
${\cal H}  \to {\cal H} $) since, using  $\chi_+ := i\psi_+$ above, one gets 
 $S(g)( i \psi_+) = ((i\psi_+ -i \overline{\psi_+})^{(g^{-1})})_+ =
 i  ((\psi_+ -\overline{\psi_+})^{(g^{-1})})_+ \neq i ((\psi_+ +\overline{\psi_+})^{(g^{-1})})_+ = i S(g)\psi_+$.
Actually, if $g\in PSL(2,\bR)$, it turns out that $(\overline{\psi_+}\circ g^{-1})_+ =0$ so that  
 $S(g) \psi_+ = (\psi_+\circ g^{-1})_+$ and $S$ is $\bC$-linear.  This  nontrivial 
result was proved  in Lemma i	\ref{lemma1}. To conclude the proof it is sufficient to show that
$S(g)= U (g)\spa\rest_{{\cal H} }$ for all $g\in PSL(2,\bR)$. To establish such an identity
we first notice that $S(g): {\cal H}  \to {\cal H} $ is a unitary representation of $PSL(2,\bR)$. The only fact non self-evident is that
$S(g)$ preserve the scalar product.
 It is however  true because, if $\chi:= i\psi_+ -i\overline{\psi_+}$, 
it holds 
$$\langle\psi_+,\psi'_+ \rangle = -i\Omega(\overline{\psi_+}, \psi'_+) = \frac{-i}{2}\left( \Omega(\psi,\psi') +i \Omega(\chi,\psi')\right)$$
now, due to (\ref{center}) we can replace the arguments $\psi_+, \psi'_+$  by
respectively $S(g)\psi_+, S(g)\psi'_+$ and the arguments $\psi,\psi', \chi$  by 
$\psi^{(g^{-1})},\psi'^{(g^{-1})}, \chi^{(g^{-1})}$ respectively,   obtaining a similar identity;  finally, since 
the action of positive-oriented  diffeomorphisms of $\bS^1$ preserves the symplectic form,  one has
$\Omega(\psi^{(g^{-1})},\psi'^{(g^{-1})}) +i \Omega(\chi^{(g^{-1})},\psi'^{(g^{-1})})= \Omega(\psi,\psi') +i \Omega(\chi,\psi')$ and thus $\langle S(g)\psi_+,S(g)\psi'_+ \rangle =\langle\psi_+,\psi'_+ \rangle$. To conclude the proof it is sufficient to notice that, by direct inspection making use of 
Stone theorem one finds \footnote{Details are very similar to those in the  corresponding part of Theorem 
2.4 in \cite{mopi4}} that, if $\psi_{nj}= \{\delta_{np}\}_{p=1,2,\ldots}\otimes u_j\in \ell^2(\bC)\otimes L^2(\Sigma,\omega_\Sigma) ={\cal H} $
$$iX \psi_{nj} = \frac{d}{dt} S(exp(tx))\psi_{nj}$$
where $X= K,S,D$ and, respectively, $x=k,s,d$ ($k,d,s$ being the basis of $sl(2,\bR)$ introduced
above). 
On the other hand the same result holds, by construction, for the representation $U \spa\rest_{{\cal H} }$
$$iX \psi_{nj} = \frac{d}{dt} U (exp(tx))\psi_{nj}\:.$$
 Since the elements $\psi_{nj}$
span a dense space of analytic vectors for $K\spa\rest_{{\cal H} }^2+ S\spa\rest_{{\cal H} }^2+D\spa\rest_{{\cal H}}^2$,
by the results by Nelson cited in the proof of theorem \ref{1},  
$S=U \spa\rest_{{\cal H}}$. Now  (\ref{center}) implies (\ref{b3}) and this concludes the proof. \\
$\Box$\\

\noindent{\em Proof of Theorem \ref{4}}. 
 (a)  Consider the closure $\cW_\zeta(\bM)$   of the $^*$-algebra of 
 in $\cW(\bM)$  spanned 
 elements  $V_\zeta(\omega):=V(\omega)e^{i\int_{\bM} \Gamma ( \zeta \omega_+ + \overline{ \zeta \omega_+})}$ with $\omega\in \cD(\bM)$. 
 Obviously the obtained  $C^*$-algebra coincides with $\cW(\bM)$ itself. On the other hand 
  its  generators  $V_\zeta(\omega)$ satisfy (V1) and (V2) and thus, by theorem 5.8.8 in \cite{BR2} 
 there is a unique  $^*$-isomorphism $\gamma_\zeta : \cW(\bM) \to \cW_\zeta(\bM) =
  \cW(\bM)$ with
 $\gamma_\zeta( V(\omega)) = V(\omega)e^{i\int_{\bM} \Gamma ( \zeta \omega_+ + \overline{ \zeta \omega_+})}$. Finally, by construction 
 $\lambda(\gamma_\zeta(V(\omega)))= \lambda_\zeta(V(\omega))$ and thus, linearity and continuity 
 imply (\ref{GNS'}). 
  Let us proof (\ref{as2}). Due to linearity and continuity, it is sufficient to show the validity of the relation 
  when restricting to elements $V_\zeta(\omega)$. In turn, 
  since $V(\omega)$ is invariant under  $g_t:= \exp(t\cD)$ and using lemma \ref{lemma1},
  the  validity of  (\ref{as2}) for those elements 
  is a consequence of the invariance of the integral $\int_{\bM}\zeta \Gamma \omega_+$ under the action of 
 $g^*_t$ on the argument $\omega_+$ which we go to prove.
 If $\cD(\bM)\ni \omega = \frac{\partial f(\theta,s)}{\partial \theta} d\theta \wedge \omega_\Sigma(s)$ and defining 
 $\theta_\pm(v)= \pm 2\tan^{-1} (e^{v})$,  direct computation yields:
 $$\int_{\bM} \zeta\Gamma \: \omega_+ = -\lim_{N\to +\infty}
 \int_{-N}^{N} dv \int_{\Sigma} \omega_\Sigma(s) \zeta(s) \left[ f_+(\theta_+(v),s) - f_+(\theta_-(v),s)\right]
 + \mbox{boundary terms}\:.$$
Using periodicity of $f_+$ in $\theta$,  boundary terms can be re-arranged into a  term
$$ \lim_{\Theta\nearrow \pi} \left[(\Theta -\pi) \ln\left(\left|\tan\frac{\Theta}{2}\right| \right)\:
\int_\Sigma\zeta(s) \frac{f_+(\Theta,s)- f_+(\pi,s)}{\Theta -\pi}\: \omega_\Sigma  \:\right]$$
and three other similar terms where $-\pi$ or $0$ replaces $\pi$. The last integral can be bounded
uniformly  in $\Theta$ using Lagrange theorem since 
$\frac{\partial f_+}{\partial \theta}$ is continuous and compactly supported. As a consequence the limit vanishes and the boundary terms can be dropped. Finally, using the fact that $v$ is the parameter of the integral curves of $\cD$ one has,
 \beq \int_{\bM} \zeta\Gamma \: g_t^*\omega_+ &=& -\lim_{N\to +\infty}
 \int_{-N}^{N} dv \int_{\Sigma} \omega_\Sigma(s)\zeta(s)
 \left[ f_+(\theta_+(v-t),s) - f_+(\theta_-(v-t),s)\right]\nonumber \\ &=& -\lim_{N\to +\infty}
 \int_{-N+t}^{N+t} dv \int_{\Sigma} \omega_\Sigma(s)\zeta(s) \left[ f_+(\theta_+(v),s) - f_+(\theta_-(v),s)\right]
 =  \int_{\bM} \zeta \Gamma \: \omega_+ \nonumber \:,\eeq
 so that the invariance of the integral functional under $\exp(t\cD)$  is evident.\\ 
  (b) Let us start from the bottom. Since $\lambda$ is invariant under the whole 
  $PSL(2,\bR)$ group, invariance (noninvariance) 
 of $\lambda_\zeta$ is equivalent to invariance (noninvariance) of the  integral functional in the right-hand side of (\ref{GNS'}).   
 Let us study that integral.
 Take $\omega(\theta, s) = \frac{\partial f(\theta)}{\partial v} h(s)d\theta \wedge \omega_\Sigma(s)$ where $s$ are coordinates
 on $\Sigma$ and the real functions $f$ and $h$ are smooth with the latter compactly supported as well.  Assume $\zeta\neq 0$ a.e.
  We can fix $h$ such that 
 $\int_\Sigma \zeta h = e^{i\alpha}$.
 In this case
 $$\int_{\bM} \Gamma(\zeta \omega_+ + \overline{\zeta \omega_+})
 = \int_{\bS^1} \Gamma(\theta) (e^{i\alpha} \frac{\partial f_+}{\partial \theta} d\theta+ c.c.)\:.$$
 As a consequence, if $\{g_t\}_{t\in \bR}$ denotes the one-parameter subgroup of
 $PSL(2,\bR)$ generated by $X= (a + b\cos \theta + c\sin \theta) \partial_\theta$,
 with $a,b,c\in \bR$,
 one has:
  $$\frac{d}{dt}|_{t=0}\int_{\bM} \Gamma(\zeta g_t^*\omega_+ + \overline{\zeta g^*_t\omega_+})
 = \int_{\bS^1} \Gamma(\theta) \left(e^{i\alpha} \frac{\partial }{\partial \theta}\left((a + b\cos \theta + c\sin \theta) \frac{\partial f_+}{\partial \theta}
 \right) d\theta+ c.c. \right)\:.$$
  The invariance of the integral implies that the left-hand must  vanish no matter the choice of $f$:
  $$\int_{\bS^1} \Gamma(\theta)  \frac{\partial }{\partial \theta}\left((a + b\cos \theta + c\sin \theta) \frac{\partial e^{i\alpha} f_+}{\partial \theta}
 \right) d\theta+ c.c.  =0\:.$$
  Using $f(\theta):= \cos(\theta-\alpha)$ one finds that it must be $a=0$ as a consequence 
  of  the identity above. Then using $f(\theta):= \cos(2\theta-\alpha)$ one finds that it must also 
  be $b=0$.  We conclude that the integral functional is invariant at most under the 
 group generated by $c\sin \theta \partial/\partial \theta = -c\cD$. On the other hand the proof of such an invariance arises directly from (\ref{as1}) and (\ref{as2}) 
 using the fact that $\lambda$ is invariant under $\alpha_t^{(\cD)}$ 
 as stated in  (c) in theorem \ref{3}.\\
 The fact that $\lambda_\zeta$ is pure (that is extremal) is an immediate consequence of (\ref{GNS'})
 using the fact that $\gamma_\zeta$ is bijective and $\lambda$ is pure. As  the $\lambda_\zeta$ are pure  their GNS 
representations are irreducible.
  Therefore the proof of the fact that $\lambda_\zeta$ and $\lambda_{\zeta'}$ are not quasiequivalent if $\zeta\neq \zeta'$ a.e.
  reduces to the proof that,  if $\zeta\neq \zeta'$ a.e.,  there is no unitary  
  transformation 
$U :  \gF_+(\cH_{\zeta}) \to \gF_+(\cH_{\zeta'})$ such that $U\hat{V}_{\zeta}(\omega) U^{-1} =
\hat{V}_{\zeta'}(\omega)$ 
 for all $\omega\in \cD(\bM)$.  
  We shall make use of the first statement in (c) which will be proved independently from 
  the following. 
   Suppose that there is such  a unitary transformation for some choice of $\zeta\neq \zeta'$.
   As a consequence one gets also the identity
o   $U\hat{V}_{\zeta}(\omega) e^{-i(\int_\bM \zeta \Gamma \omega_+ + c.c.)}
    U^{-1} =
\hat{V}_{\zeta'}(\omega)e^{-i(\int_\bM \zeta \Gamma \omega_+ +c.c.)}$. That is, re-defining 
$\zeta'-\zeta \to \zeta \neq 0$, one has
$U e^{i\overline{\hat{\phi}_\zeta(\omega)}} U^{\dagger} =
e^{i\overline{\hat{\phi}_0(\omega)}}$
where we have also identified the one-particle Hilbert spaces ${\cal H}_0$ and $\cH_\zeta$
with the one-particle space $\cH$ of the GNS representation of $\lambda$
(and thus the Fock spaces). Via Stone theorem (using above $\omega= t\omega$ and $t\in \bR$) one gets
$U \overline{\hat{\phi}_\zeta(\omega)} =
\overline{\hat{\phi}_0(\omega)}U$, that is $i\:U \overline{a(\overline{\psi_+}) - a^\dagger({\psi_+})} + (\int_{\bM} \zeta 
\Gamma \epsilon_{\psi_+} + c.c.) U
=  i\:\overline{a(\overline{\psi_+}) - a^\dagger({\psi_+})} U$
where $\psi_+ = E\omega_+$ according with (b) in proposition 2.1. Using the analogous relation for 
$\psi' := i\psi_+ -i \overline{\psi_+}$ one gets in the end 
$$U \spa\left[\overline{a(\overline{\psi_+}) - a^\dagger({\psi_+})}
+  \overline{a(\overline{\psi_+}) + a^\dagger({\psi_+})}\right]
 - \left(\spa 4i\spa \int_{\bM}\spa\spa\overline{\zeta} \Gamma \epsilon_{\overline{\psi_+}}\spa\right)\spa U
=  \left[\overline{a(\overline{\psi_+}) - a^\dagger({\psi_+})} +
 \overline{a(\overline{\psi_+}) + a^\dagger({\psi_+})}\right]\spa U.$$
 Applying both sides to the vacuum state  $\Psi_\zeta$ and  computing the scalar 
 product of the resulting vectors with  $\Psi_\zeta$ itself,
 the identity above implies that
 $$ - \left(2i\int_\bM \overline{\zeta} \Gamma \epsilon_{\overline{\psi_+}}\right) \langle \Psi_\zeta, 
U\Psi_\zeta \rangle = \langle a^\dagger({\psi_+}) \Psi_\zeta,U\Psi_\zeta\rangle\:.$$
 If $\{\psi_{+m}\}_{m\in \bN'}$ is a Hilbert base of $\cH_\zeta$, iteration of the procedure 
sketched above produces
\beq  \langle \Psi_\zeta, U\Psi_\zeta \rangle \prod_{n} \frac{\lambda^{N_m}_m}{\sqrt{N_m!}} = \langle N_1,N_2,\ldots,N_m, \ldots | U\Psi_\zeta\rangle \label{OC} \eeq
for any vector with finite number of particles $|N_1,N_2,\ldots,N_m, \ldots\rangle$, $N_m$
being the occupation number of the state $\psi_{+m}$ and where 
$\lambda_m := -2i\int_\bM \overline{\zeta}\Gamma \epsilon_{\overline{\psi_{+m}}}$. 
It must be $ \langle \Psi_\zeta, U\Psi_\zeta \rangle \neq 0$, otherwise
all components of $U\Psi_\zeta$ would vanish producing $U\Psi_\zeta=0$
which is impossible since $U$ is unitary. Conversely, as $||\Psi_\zeta||^2=1$, it
must hold $||U \Psi_\zeta||^2=1$. This identity can be expanded with the basis of  states $|N_1,N_2,\ldots,N_m, \ldots\rangle$ and a straightforward computations which employs  (\ref{OC}) 
produces
\beq ||U\Psi_\zeta||^2=   |\langle \Psi_\zeta, U\Psi_\zeta \rangle|^2  \exp\left({\sum_{m=1}^{+\infty} |\lambda_m|^2}\right)\:. \label{series}\eeq
The series can explicitly be computed using a basis
$\psi_{(n,j)}(\theta,s) = u_j(s)\frac{e^{-in\theta}}{\sqrt{4\pi n}}$ where $u_j$
is any  basis of $L^2(\Sigma, \omega_\Sigma)$ made of compactly supported real smooth functions\footnote{The space 
$\cC$ of 
smooth compactly supported functions on $\Sigma$ is dense in $L^2(\Sigma,\omega_\Sigma)$. As the latter is separable
$\cC$ contains a countable subset $\cC'$ still dense in $L^2(\Sigma,\omega_\Sigma)$. In turn one may extract 
from $\cC'$ a subset $\cC''$ of linearly independent elements which span the same dense space as $\cC'$. Usual 
orthonormalization procedure applied to $\cC''$ gives a Hilbert basis for $L^2(\Sigma,\omega_\Sigma)$ made
of smooth compactly supported functions. Proceeding as in footnote 4 one gets the wanted basis of $u_j$.}.
 In that case $\int_\Sigma \overline{\zeta} u_j \omega_\Sigma \neq 0$ for some $j=j_0$ (otherwise the  function $\zeta$
 on $\Sigma$ would have $L^2(\Sigma,\omega_\Sigma)$-norm zero).
  One finds
$|\lambda_{2n+1,j_0}|^2 = C|\int_\Sigma \overline{\zeta} u_{j_0} \omega_\Sigma|^2 (2n+1)^{-1}$ with $C>0$ so that the series in (\ref{series}) diverges
and the found contradiction shows that $U$ cannot exist.  \\
  (c) By direct inspection one finds that the operators $V_\zeta(\omega)$ enjoy (V1) and (V2). Therefore,
(theorem 5.2.8, in \cite{BR2})  the $C^*$-algebra $\hat{\cW}_\zeta(\bM)$ given by the closure of the  $^*$-algebra generated by  $V_\zeta(\omega)$ is a representation of Weyl algebra and 
  there is a  $^*$-algebra isomorphism of $C^*$ algebras, $\Pi_\zeta: \cW(\bM) \to \hat{\cW}_\zeta(\bM)$ which satisfies (\ref{b}). The vacuum vector of $\gH_\zeta= \gF_+(\cH_\zeta)$
  is cyclic with respect to $\Pi_\zeta$ because $\hat{\cW}_\zeta(\bM)\Psi_\zeta$ is the same 
  space as the dense space (see A.1) spanned by vectors $e^{i\overline{\hat{\phi}(\omega_1)}}\cdots e^{i\overline{\hat{\phi}(\omega_n)}}\Psi_\zeta$,
  $n=,1,2,\ldots$. Finally it holds
  $$\lambda_\zeta(V(\omega)) =  \lambda(V(\omega)) e^{i(\int_{\bM}\zeta\Gamma\omega_+ +c.c.)} =  \langle\Psi_\zeta,   
  e^{i\overline{\hat{\phi}(\omega)}} \Psi_\zeta \rangle e^{i(\int_{\bM}\zeta\Gamma\omega_{+} +c.c)}  =
   \langle\Psi_\zeta, e^{i(\overline{\hat{\phi}(\omega) + \int_{\bM}\zeta\Gamma\omega_{+} + c.c.})}\Psi_\zeta\rangle$$
   $$ =
 \langle \Psi_\zeta, \hat{V}_\zeta(\omega) \Psi_\zeta\rangle\:,$$ that is
 $\lambda_\zeta(V(\omega)) =  \langle \Psi_\zeta, \Pi_\zeta(V(\omega)) \Psi_\zeta\rangle$.
  By linearity and  continuity this relation extends to the whole algebras:
 $\lambda_\zeta(w) =  \langle \Psi_\zeta, \Pi_\zeta(w) \Psi_\zeta\rangle$, $w\in \cW(\bM)$.
 We conclude that $(\gF_+(\cH_\zeta),\Pi_\zeta, \Psi_\zeta)$ is the (unique, up to unitary transformations)  GNS triple for $\lambda_\zeta$. \\
 (d) Let us denote by $\{g_t\}_{t\in \bR}$ the one-parameter group of M\"obius transformations 
 generated by $\cD$.
 The statements (a) and (b) in theorem \ref{2}  imply that if $D$ is defined as 
  $(1/2i): \spa\Omega(\hat{\phi}_0,\cD(\hat{\phi_0}))\spa:$ then
 $e^{itD} e^{i\overline{\hat{\phi}_0(\omega)}} e^{-itD} = e^{i\overline{\hat{\phi}_0(g_t^{-1*}\omega)}}$.
 Since   $\int_{\bM} \zeta\Gamma\omega_+ + c.c.$ is invariant under the action of $g_t$ on $\omega$ 
as seen in the proof of (a), we  have also   $$e^{itD} e^{i\overline{\hat{\phi}_0(\omega)}} 
 e^{i(\int_{\bM} \zeta \Gamma\omega_+ +c.c.)}
 e^{-itD} = e^{i\overline{\hat{\phi}_0(g_t^{-1*}\omega)}}
 e^{i(\int_{\bM} \zeta \Gamma g_t^{-1*}\omega +c.c.)}$$ that can be rewritten as $e^{itD} 
\hat{V}_\zeta(\omega) e^{-itD} = \hat{V}_\zeta(\omega^{(g_t^{-1})})$  and thus extends to the whole Weyl algebra
proving (\ref{reduced}). $\Box$\\

\section*{Acknowledgments.}
Concerning the part of this work due to N.P., it has been funded by Provincia Autonoma di Trento within the
postdoctoral project FQLA, Rif. 2003-S116-00047 Reg. delib. n. 3479 \& allegato B.

\end{document}